\documentclass[table]{article} 

\usepackage[preprint]{neurips_2025}

\usepackage[utf8]{inputenc} 
\usepackage[T1]{fontenc}    
\usepackage{hyperref}       
\usepackage{url}            
\usepackage{booktabs}       
\usepackage{graphicx}       
\usepackage{tabularx}       
\usepackage{amsfonts}       
\usepackage{nicefrac}       
\usepackage{microtype}      
\usepackage{xcolor}         
\usepackage{multirow}
\usepackage{makecell}
\usepackage{ragged2e}      
\usepackage{wrapfig}       
\definecolor{DarkGreen}{RGB}{125, 192, 73}

\usepackage{pifont}
\usepackage{subcaption} 
\usepackage{siunitx}

\usepackage{verbatim}
\usepackage{enumitem}
\usepackage[most]{tcolorbox} 
\usepackage{float}
\usepackage{xspace}
\tcbset{
  aibox/.style={
    width=474.18663pt,
    top=10pt,
    colback=white,
    colframe=black,
    colbacktitle=black,
    enhanced,
    center,
    attach boxed title to top left={yshift=-0.1in,xshift=0.15in},
    boxed title style={boxrule=0pt,colframe=white,},
  }
}
\newtcolorbox{AIbox}[2][]{aibox,title=#2,#1}

\title{FusionAudio-1.2M: Towards Fine-grained Audio Captioning with Multimodal Contextual Fusion}

\author{%
  Shunian Chen\textsuperscript{1} \thanks{Equal contribution.} \quad
  Xinyuan Xie\textsuperscript{1,2} \footnotemark[1] \quad
  Zheshu Chen\textsuperscript{1} \footnotemark[1] \quad  \\
  Liyan Zhao\textsuperscript{1} \quad 
  Owen Lee\textsuperscript{1} \quad
  Zhan Su\textsuperscript{1} \quad
  Qilin Sun\textsuperscript{1} \quad
  Benyou Wang\textsuperscript{1} \thanks{Corresponding author} \\[1.5ex] 
  \textsuperscript{1}The Chinese University of Hong Kong, Shenzhen \\
  \textsuperscript{2}South China University of Technology \\[0.5ex] 
\texttt{wangbenyou@cuhk.edu.cn}
}

\begin{document}

\maketitle

\begin{abstract}
High-quality, large-scale audio captioning is crucial for advancing audio understanding, yet current automated methods often generate captions that lack fine-grained detail and contextual accuracy, primarily due to their reliance on limited unimodal or superficial multimodal information. Drawing inspiration from human auditory perception, which adeptly integrates cross-modal cues and performs sophisticated auditory scene analysis, we introduce a novel two-stage automated pipeline. This pipeline first employs specialized pretrained models to extract diverse contextual cues (e.g., speech, music, general sounds, and visual information from associated video). A large language model (LLM) then synthesizes these rich, multimodal inputs to generate detailed and context-aware audio captions. Key contributions of this work include: (1) the proposed scalable method for fine-grained audio caption generation; (2) FusionAudio, a new large-scale dataset comprising 1.2 million such detailed captions, combined with 6 million QA pairs; and (3) enhanced audio models developed using FusionAudio, specifically a CLAP-based audio encoder with superior audio-text alignment and instruction following.
This paper paves the way for more nuanced and accurate automated understanding of complex audio environments. Code and data can be found in \url{https://github.com/satsuki2486441738/FusionAudio}.
\end{abstract}

\section{Introduction}

The advancement of models like CLAP~\cite{laionclap2023} for audio retrieval, and GAMA~\cite{ghosh-etal-2024-gama} or Qwen2-Audio~\cite{Qwen2-Audio} for broader audio understanding, heavily relies on large-scale, high-quality audio captioning datasets. 
Audio captioning has primarily followed two trajectories. \textit{Manual annotation} ~\cite{drossos2019clothoaudiocaptioningdataset,kim-etal-2019-audiocaps} offers high quality but lacks scalability due to high labor costs. In contrast, \textit{automated methods}~\cite{laion-audio-630k,mei2023wavcaps} often use sparse metadata like text labels or tags to assist annotation, while others ~\cite{bai2024audiosetcapsenrichedaudiocaptiondataset, sun2024autoacdlargescaledatasetaudiolanguage, yuan2025soundvecapsimprovingaudiogeneration} leverage basic multimodal cues. These automated approaches, however, typically rely on limited textual or superficial  information, failing to capture rich  details (e.g., multimodal contextual details). This results in captions that lack fine-grained details and are prone to hallucinations~\cite{yang2024airbenchbenchmarkinglargeaudiolanguage}, hindering nuanced audio interpretation.

\begin{wrapfigure}{r}{0.33\columnwidth}
    \vspace{-20pt}
    \centering
    \includegraphics[width=\linewidth]{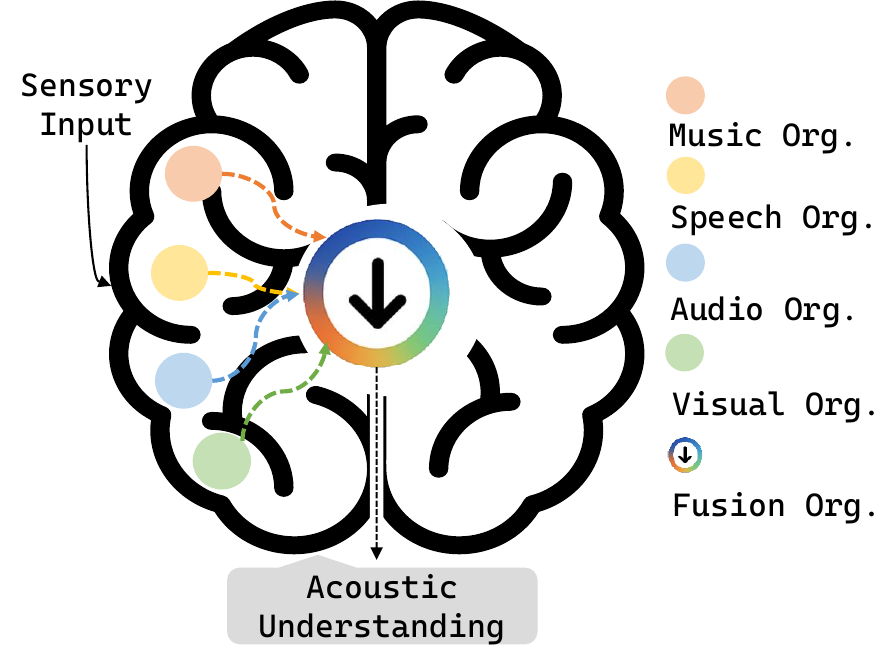}
    \caption{Human auditory perception integrates multisensory cues.}
    \label{fig:analogy_comparison_wrap}
    \vspace{-10pt}
\end{wrapfigure}

Addressing this gap necessitates a paradigm shift. We turn to human auditory perception for inspiration (Figure~\ref{fig:analogy_comparison_wrap}). Human auditory understanding leverages sophisticated strategies at two complementary levels. \textbf{Firstly}, humans adeptly integrate cross-modal cues---visual information, for instance, aids speech intelligibility~\cite{sumby1954visual} and sound identification~\cite{Kayser2010, ERNST2004162}. \textbf{Secondly}, auditory scene analysis (ASA)~\cite{bregman1990auditory} allows the auditory system to parse complex soundscapes into distinct streams like speech, music, and ambient sounds based on temporal-spectral regularities~\cite{Shamma2011}. These sophisticated biological mechanisms offer a compelling blueprint for enhancing automated audio captioning. The impact of this multimodal integration is demonstrated in Table~\ref{tab:comparison}. Current systems, often processing audio in isolation, can misinterpret sounds (e.g., a stationary motorcycle as a moving scooter) or hallucinate details. In contrast, FusionAudio-1.2M leverages comprehensive audiovisual cues to produce more accurate and contextually rich descriptions.

Inspired by these principles, we introduce a two-stage pipeline for enhanced automated audio captioning. First, specialized pretrained models extract diverse contextual cues: an Automatic Speech Recognition (ASR) model~\cite{radford2022robustspeechrecognitionlargescale} for speech, a music understanding model~\cite{zhao2024openmu} for musical attributes, an audio understanding model~\cite{ghosh-etal-2024-gama} for general sounds, and a visual model~\cite{Qwen2.5-VL} for video information. Second, a large language model (LLM)~\cite{qwq32b} acts as an integration engine, synthesizing these multimodal cues to generate fine-grained audio captions. This synthesis of rich, cross-modal context by an LLM aims to improve detail and accuracy, addressing prior limitations.

\begin{table}[t] 
    \centering
    \small
    \caption{Comparison of generated captions for a sample audio clip with associated visual context. Hallucinations in prior work are highlighted in \textcolor{red}{red}. Improvements from our multimodal approach, FusionAudio, are highlighted in \textcolor{DarkGreen}{green}, demonstrating enhanced accuracy and detail by leveraging visual and comprehensive auditory cues.}
    \begin{tabularx}{\textwidth}{@{} l >{\RaggedRight\arraybackslash}X @{}}
        \multicolumn{2}{c}{\includegraphics[width=0.6\linewidth, keepaspectratio]{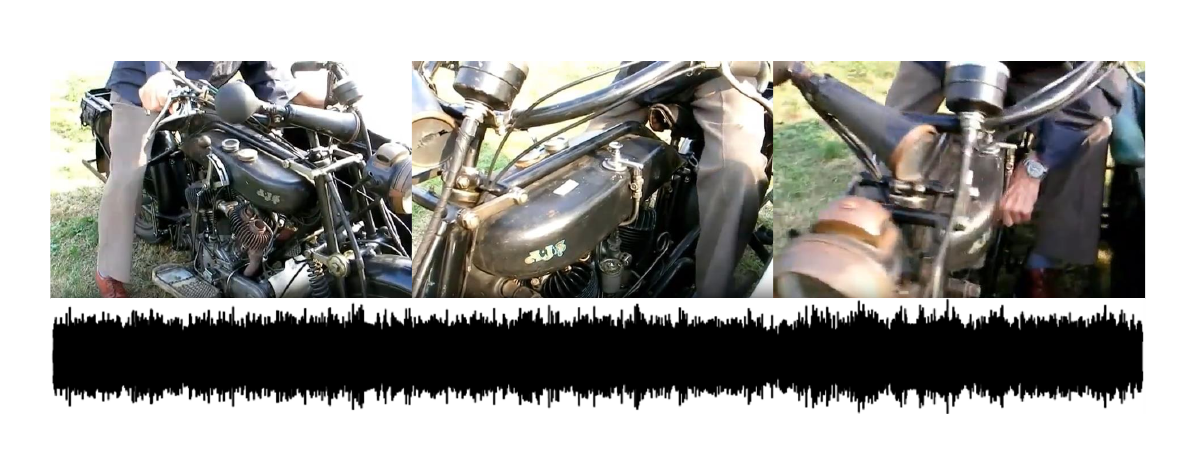}} \\
        \addlinespace
        \toprule
        \textbf{Method} & \textbf{Audio Caption} \\
        \midrule
        \textbf{GAMA (Baseline)} & The audio is dominated by the sound of a motor vehicle engine and intermittent male speech, with wind noise. \\
        \textbf{AudioSetCaps} & A male and \textcolor{red}{female} engage in conversation, their voices audible against a backdrop of \textcolor{red}{ambient noise}. The discussion is neutral in tone and does not involve any identifiable objects or language. \\
        \addlinespace
        \textbf{Auto-ACD}       & A man speaks while a vehicle \textcolor{red}{moves in the distance}, possibly on a motor scooter, in an \textcolor{red}{engine room}. \\
        \addlinespace
        \textbf{Sound-VECaps}    & A man is speaking and a vintage motorcycle \textcolor{red}{with a large headlamp, round fuel tank, and sidecar is parked on grass}, with the sound of the engine and the man's voice filling the air, while \textcolor{red}{a vehicle passes by} in the background. \\
        \addlinespace
        \textbf{FusionAudio-1.2M (Ours)} & \textcolor{DarkGreen}{Continuous} motor vehicle engine noise is prominently featured, accompanied by intermittent male speech with \textcolor{DarkGreen}{a positive or confirming tone}. Wind sounds suggest an \textcolor{DarkGreen}{outdoor environment}, with the engine's sustained roar maintaining a steady volume throughout the recording. \\ 
        \bottomrule
    \end{tabularx}
    \label{tab:comparison}
\end{table}


Our contributions are:
\begin{itemize}
    \item Automated fine-grained audio captioning: A pipeline using specialized unimodal models to extract diverse contextual cues, synthesized by an LLM to generate detailed, scalable captions.
    \item FusionAudio-1.2M dataset: A large-scale dataset of 1.2M fine-grained audio captions to advance audio research.
    \item Multimodal cue-enhanced audio models:  A CLAP-based audio encoder with improved audio-text alignment, and an instruction-tuned MLLM with stronger audio comprehension and instruction-following.
\end{itemize}

\section{Related Works}

\subsection{Audio Language Learning}
The field of audio-language models has seen significant advancements in recent years, with researchers focusing on developing models that can effectively process, understand, and reason about sounds using natural language as a supervision signal. Early works like CLAP Learning Audio Concepts from Natural Language Supervision~\cite{elizalde2023clap} laid the foundation for contrastive learning approaches in audio-language pre-training. Subsequent studies have explored various pre-training objectives, including generative and discriminative methods, to enhance audio representation learning and cross-modal alignment. For instance, CTAL~\cite{li2021ctal} and FLAP~\cite{yeh2023flap} have investigated masked language modeling and cross-attention based masked acoustic modeling for joint representation learning of audio and language modalities. Multi-task learning approaches have also gained attention, with models like UniAudio~\cite{tian2023uniaudio} and SpeechX~\cite{wang2024speechx} demonstrating the potential of unifying diverse audio tasks under a single framework. Furthermore, the integration of large language models (LLMs) with audio processing has opened new avenues for creating more powerful and human-like audio understanding systems. Recent works such as Pengi~\cite{deshmukh2023pengi}, Qwen-audio ~\cite{chu2023qwen}, and Audio Flamingo~\cite{kong2024audio} have shown impressive capabilities in handling complex audio tasks through instruction tuning and in-context learning. These advancements highlight the growing importance of audio-language models in bridging the gap between auditory information and language understanding, and their potential for real-world applications.

\begin{table}[t] 
\centering
\caption{Comparison of open-source audio caption datasets. }
\resizebox{\textwidth}{!}{  
\begin{tabular}{lcccccccc}
\toprule
\textbf{Name} & \textbf{Year} & \textbf{\# of Audio/QA} & \textbf{Avg. Dur (s)} & \textbf{Avg. Text Len} & \textbf{Visual} & \textbf{Music} & \textbf{Speech} & \textbf{Integration}\\
\midrule
AudioCaps~\cite{kim-etal-2019-audiocaps}         & 2019 & 46k/46k   & 10.00     & 9.03      & \ding{55} & \ding{55} & \ding{55} & \ding{55} \\
Clotho~\cite{drossos2019clothoaudiocaptioningdataset}         & 2019 & 5k/5k   & 22.50    & 11.00      & \ding{55} & \ding{55} & \ding{55} & \ding{55} \\
LAION-Audio-630K~\cite{laion-audio-630k}         & 2022 & 630k/630k   & 24.58     &   7.30    & \ding{55} & \ding{55} & \ding{55} & \ding{55} \\
WavCaps~\cite{mei2023wavcaps}         & 2024 & 403k/403k   & 67.59     & 7.80      & \ding{55} & \ding{55} & \ding{55} & \ding{55} \\
AudioSetCaps~\cite{bai2024audiosetcapsenrichedaudiocaptiondataset} & 2024 & 1.9M/1.9M  &   N/A   & 28.00        & \ding{55} & \ding{55}  & \ding{55} & \ding{55} \\
Auto-ACD~\cite{sun2024autoacdlargescaledatasetaudiolanguage} & 2024 & 1.5M/1.5M  &  10.00     & 18.10        & \ding{51}  & \ding{55}  & \ding{55} & \ding{51} \\
CompA-R~\cite{ghosh-etal-2024-gama} & 2024 & 62k/200k  &  9.93     & 18.00     & \ding{51}  & \ding{55}  & \ding{55} & \ding{51} \\
\rowcolor[gray]{0.9}
\textbf{FusionAudio-1.2M (Ours)} & \textbf{2025} & \textbf{1.2M/6M} & \textbf{10.00} & \textbf{47.18} & \textbf{\ding{51}} & \textbf{\ding{51}} & \textbf{\ding{51}} & \textbf{\ding{51}} \\
\bottomrule

\end{tabular}
}
\label{tab:datasets_statistics}
\end{table}

\subsection{Audio Captioning}
Early audio captioning research relied on \textit{manually annotated} datasets like AudioCaps~\cite{kim-etal-2019-audiocaps} and Clotho~\cite{drossos2019clothoaudiocaptioningdataset}, which provided high-quality descriptions but were inherently limited in scale. To address this, the field increasingly adopted \textbf{automated and weakly-supervised methods}. These leverage large-scale web-sourced audio with associated sparse metadata (e.g., WavCaps~\cite{mei2023wavcaps}, LAION-Audio-630K~\cite{laion-audio-630k}), employ existing textual tags to guide generation, or incorporate basic multimodal cues from loosely associated content~\cite{bai2024audiosetcapsenrichedaudiocaptiondataset, sun2024autoacdlargescaledatasetaudiolanguage, yuan2025soundvecapsimprovingaudiogeneration}. While significantly improving scalability, these automated techniques typically yield captions lacking the fine-grained detail and rich contextual understanding characteristic of human annotations or, as our work posits, achievable through more sophisticated, deeply integrated multimodal information processing. As shown in Table~\ref{tab:datasets_statistics}, Our FusionAudio-1.2M dataset offers  finer-grained captions  than existing audio caption dataset, see the average text length.

\section{Method: Fine-grained Audio Caption with Multimodal Contextual Fusion} 
\label{sec:method}

\begin{figure}[h]
    \centering
    \includegraphics[width=\linewidth]{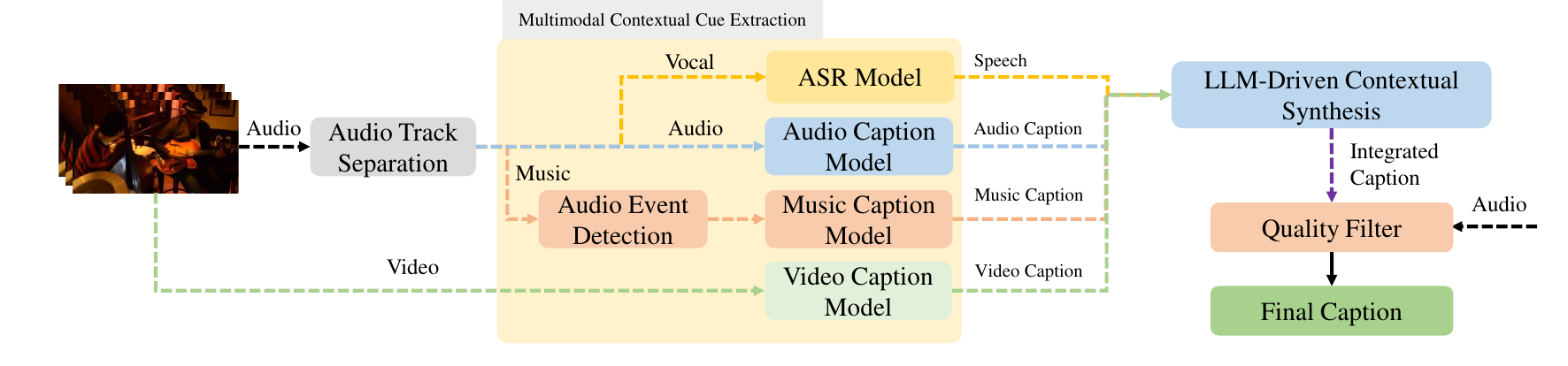}
    \caption{Overview of our proposed multimodal audio captioning pipeline. The process involves initial vocal separation, followed by a two-stage approach: multimodal contextual cue extraction and LLM-driven contextual synthesis.} 
    \label{fig:pipeline}
\end{figure}

\subsection{Automated Captioning Pipeline}
We introduce a two-stage pipeline, illustrated in Figure~\ref{fig:pipeline}, designed to generate fine-grained audio captions: (1) Multimodal Contextual Cue Extraction using specialized expert models, and (2) LLM-Driven Contextual Synthesis to integrate these diverse cues into a coherent caption. An initial pre-processing step is performed to enhance audio quality.

\paragraph{Pre-processing: Audio Track Separation.} 
To enhance the quality and specificity of downstream analyses, particularly for speech and distinct background sounds, we first apply a source separation technique. We employ the Demucs model~\cite{rouard2022hybrid} to isolate the vocal track from the non-vocal components (e.g., music, environmental sounds) within the audio stream.

\paragraph{Stage 1: Multimodal Contextual Cue Extraction.}
This stage leverages a suite of specialized models to extract diverse, complementary information streams relevant to the auditory scene. The prompts used for these models can be found in Appendix~\ref{app:prompts}.
\begin{itemize}
    \item \textbf{General Audio Events:} To capture overall acoustic scene characteristics, we utilize GAMA~\cite{ghosh-etal-2024-gama} to generate descriptive captions focusing on sound events and environments.
    \item \textbf{Speech Content:} The separated vocal stream is transcribed using the Whisper model~\cite{radford2022robustspeechrecognitionlargescale}.
    \item \textbf{Music Characteristics:} For clips potentially containing music, we first employ YamNet~\cite{yamnet2025} 
    as a classifier to confirm the presence of music, mitigating hallucination risk on non-musical segments. If music is detected, OpenMu~\cite{zhao2024openmu} is used to extract details regarding genre, instrumentation, tempo, and mood.
    \item \textbf{Visually-Grounded Context:} We utilize the Qwen2.5-VL-72B vision-language model~\cite{Qwen2.5-VL} 
    to extract visual information from the video stream. This approach yields a detailed, timestamped visual record, providing visual context that aids in grounding physical events. 
\end{itemize}

\paragraph{Stage 2: LLM-Driven Contextual Synthesis.}
The extracted information streams serve as input to the synthesis model, QwQ-32B~\cite{qwq32b}. 
The LLM acts as a central integration engine. It is prompted to: (a) synthesize the multimodal inputs coherently, (b) resolve potential redundancies or minor inconsistencies across the different expert outputs, (c) infer relationships and context implied by the combined information, and (d) generate a final, fine-grained audio caption that reflects a comprehensive understanding of the auditory scene enriched by multimodal context.

\subsection{Data Source}
We utilize the AudioSet dataset~\cite{7952261} as the primary source material. AudioSet provides over 2 million 10-second YouTube video clips, each weakly annotated with audio event labels. We downloaded the corresponding audio and video streams for processing through our pipeline.

\subsection{Data Quality Assurance}

To ensure the quality and reliability of the automatically generated captions, we implement a multi-faceted quality assurance protocol. This process involved both manual verification on a sample of the data and scalable automated filtering (described subsequently) to curate the final FusionAudio-1.2M dataset.

\paragraph{Manual Verification.} 
To establish a benchmark for caption quality, we randomly sampled 300 generated captions for human evaluation. Trained annotators assessed each caption based on two criteria:
\textbf{(1) Detailness:} Rated on a 3-point scale, a higher score means more details, evaluating the richness and specificity of the information conveyed.
    \textbf{(2) Hallucination:} Rated on a 5-point scale, a higher score means less hallucination, assessing the factual accuracy of the caption against the audio-visual content. A score of $\le 2$ was considered indicative of notable hallucination.
The detailed annotation guidelines and scoring rubrics are provided in Appendix~\ref{app:human_eval}.

As shown in Table~\ref{tab:manual_eval_results}, the manually evaluated sample achieved a mean detailness score of 2.55 (out of 3). For hallucination, the average score given by human evaluator is 3.74, with 7\% of the evaluated captions received a score of 2 or less, indicating a low prevalence of significant inaccuracies in this sample. The inter-annotator agreement, calculated using the exact match rate, was 0.67 for detailness and 0.79 for hallucination. These rates suggest moderate agreement between annotators, which is reasonable given the subjective nature of fine-grained caption quality assessment. The full distribution of scores for both metrics can also be found in Appendix~\ref{app:human_eval}.

\begin{wraptable}{r}{0.5\textwidth} 
  \centering
  \scriptsize
  \caption{Manual Verification Results. Detailness is rated 1-3 (higher is better). Hallucination is rated 1-5 (higher is better; $\le 2$ indicates notable hallucination). IAA is measured using the exact matching, before which hallucination score has been converted to 1 (score $\le$ 2) or 0 (score > 2).} 
  \label{tab:manual_eval_results} 
  \begin{tabular}{@{}cc|cc@{}}
    \toprule
    \multicolumn{2}{c}{\textbf{Caption Content Quality}}
      & \multicolumn{2}{c}{\textbf{Inter-Annotator Agreement}} \\
    \cmidrule(lr){1-2} \cmidrule(lr){3-4}
    \textbf{Detailness} & \textbf{Hallucination}
      & \textbf{Detailness} & \textbf{Hallucination} \\
    \midrule
    2.55 & 3.74 & 0.67 & 0.79 \\
    \bottomrule
  \end{tabular}
\end{wraptable}

\paragraph{Automatic Filtering}
To scale quality assessment to the entire dataset, we leveraged the CLAP model to automatically filter low-quality captions. We computed cosine similarity between CLAP-generated audio and caption embeddings as a quality indicator.
Based on our human evaluation, we categorized hallucination scores $\leq 2$ as the positive class (captions to discard) and scores $> 2$ as the negative class (captions to retain). We then evaluated various cosine similarity thresholds using the $F\textsubscript{1.05}$ score, which slightly emphasizes recall to prioritize removing hallucinated content (see Appendix~\ref{app:f_score} for computation details).


A threshold of 0.08 achieved optimal alignment with human judgments, with the exact match rate being 88.3\%. This threshold resulted in a 7.3\% filter rate and balanced false positives and negatives. We applied this validated threshold to filter the entire caption set, yielding the final 1.2 million high-quality captions in the FusionAudio-1.2M dataset.

\section{The Resulted Dataset: FusionAudio-1.2M}
\label{sec:data_stat}
\subsection{Quantitative Analysis}

Table~\ref{tab:datasets_statistics} compares our proposed dataset with other publicly available datasets. FusionAudio-1.2M distinguishes itself through its large scale, longer caption length, and integration of multiple modalities.

\begin{figure}[t] 
    \centering
    \mbox{
        \begin{subfigure}[b]{0.4\linewidth} 
            \centering
            \includegraphics[width=\linewidth]{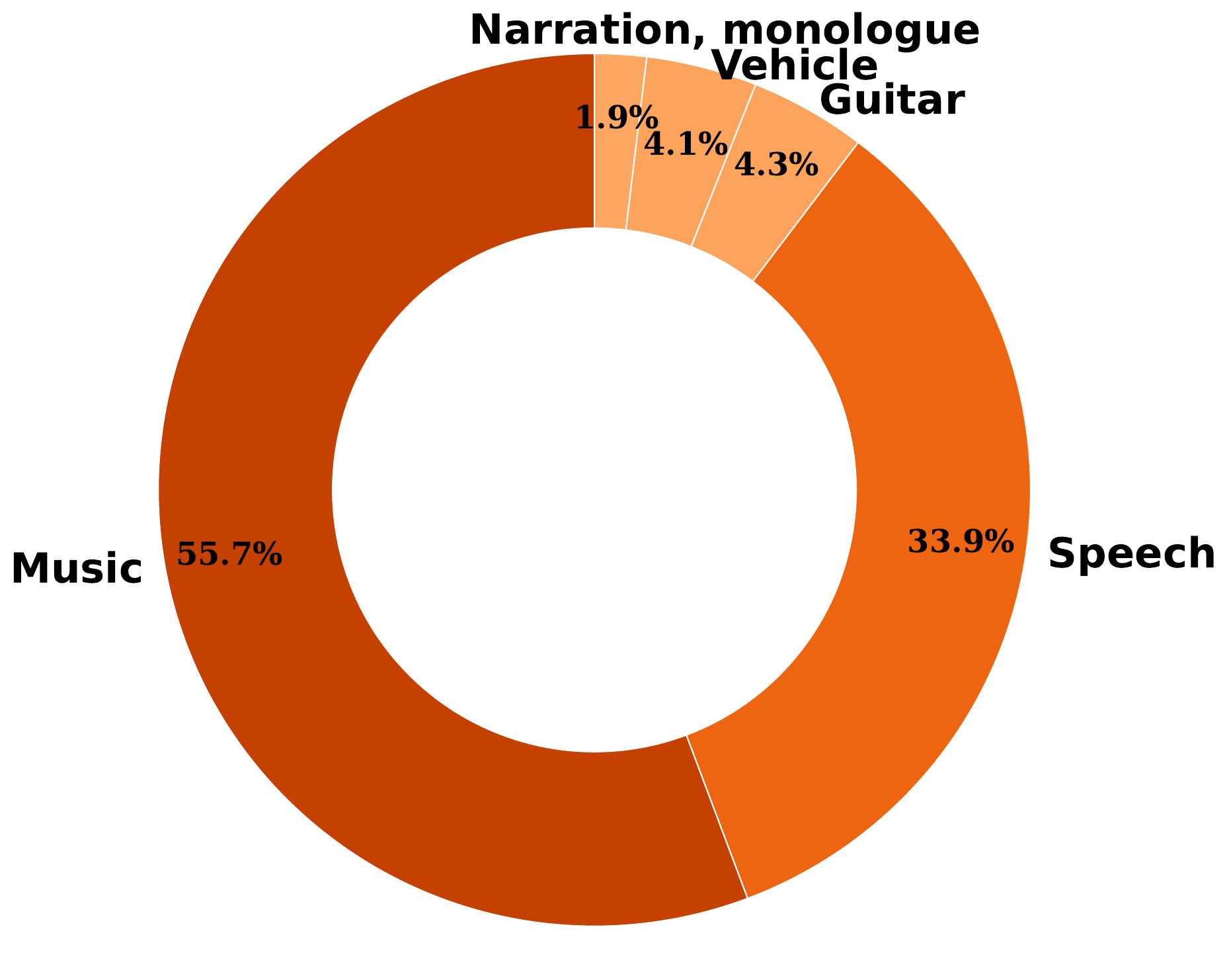}
            \caption{Top 5 Audio Labels Distribution}
            \label{fig:pie} 
        \end{subfigure}
        \hfill
        \begin{subfigure}[b]{0.6\linewidth} 
            \centering
            \begin{subfigure}[b]{0.49\linewidth}
                \includegraphics[width=\linewidth]{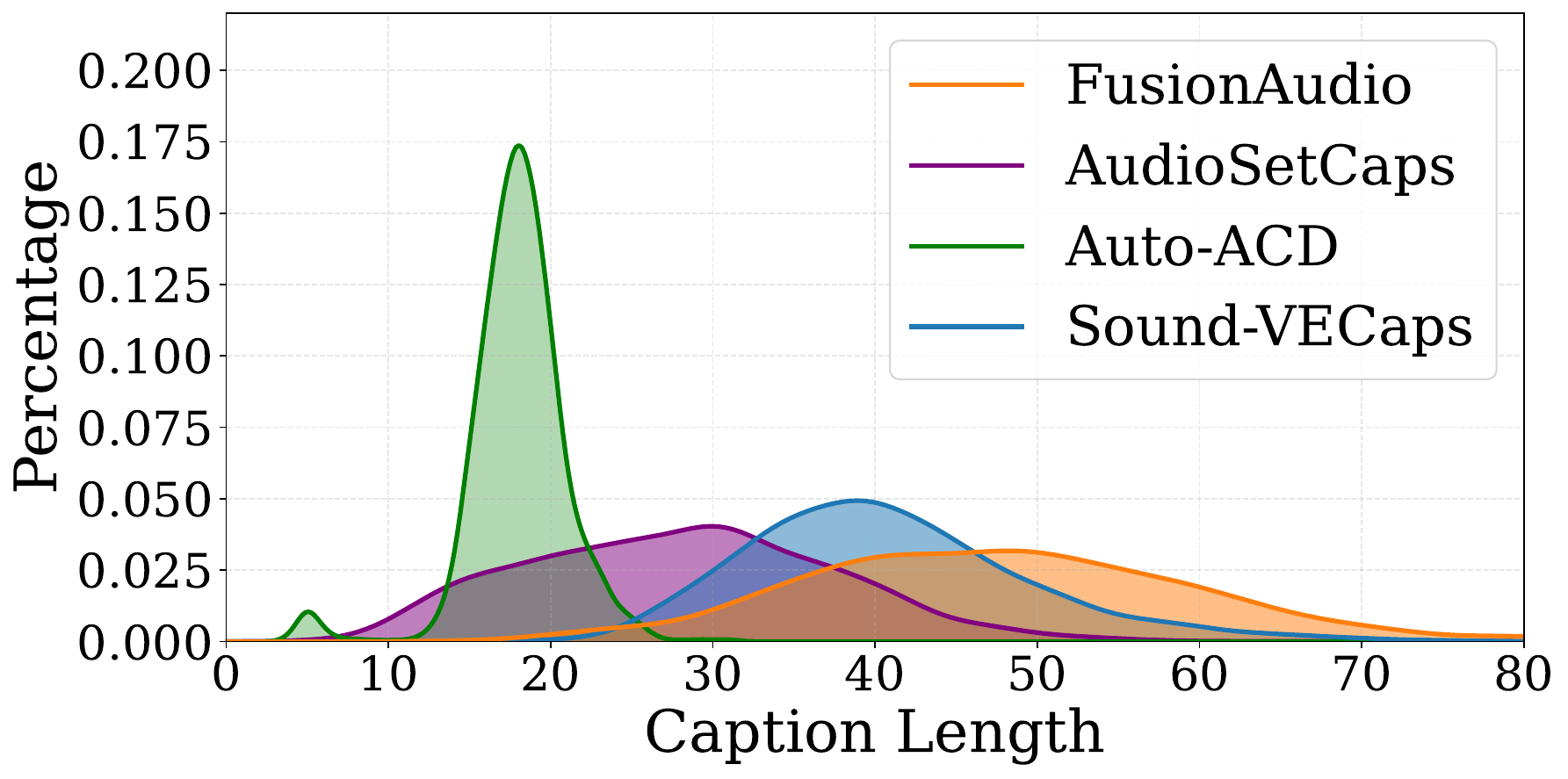}
                \footnotesize
                \caption{Caption Length Distribution}
                \label{fig:caption_length} 
            \end{subfigure}%
            \begin{subfigure}[b]{0.49\linewidth}
                \includegraphics[width=\linewidth]{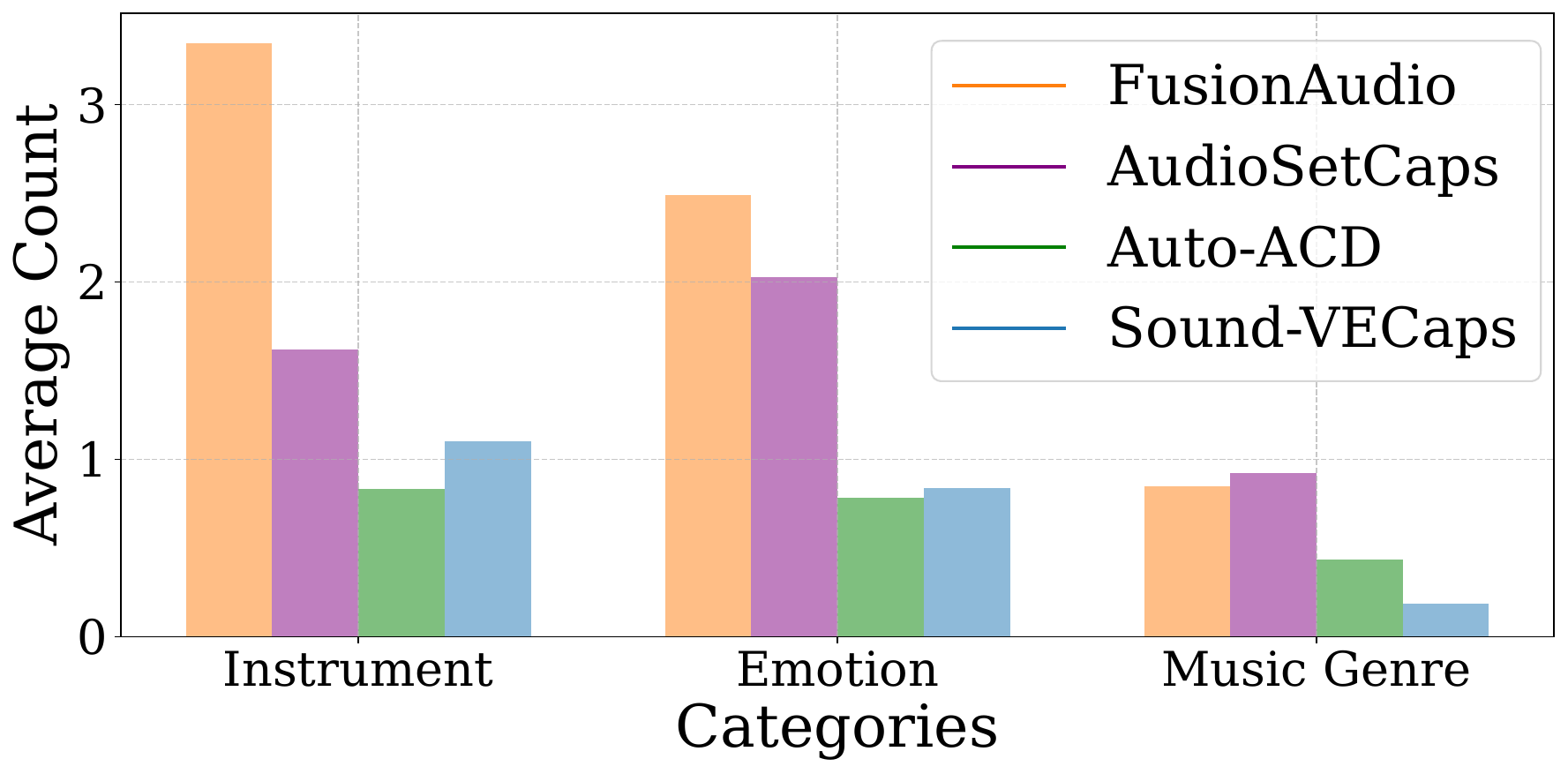}
                \caption{Diversity of Object Types} 
                \label{fig:obj_type} 
            \end{subfigure}
            
            \vspace{0.5cm}
            \begin{subfigure}[b]{0.49\linewidth}
                \includegraphics[width=\linewidth]{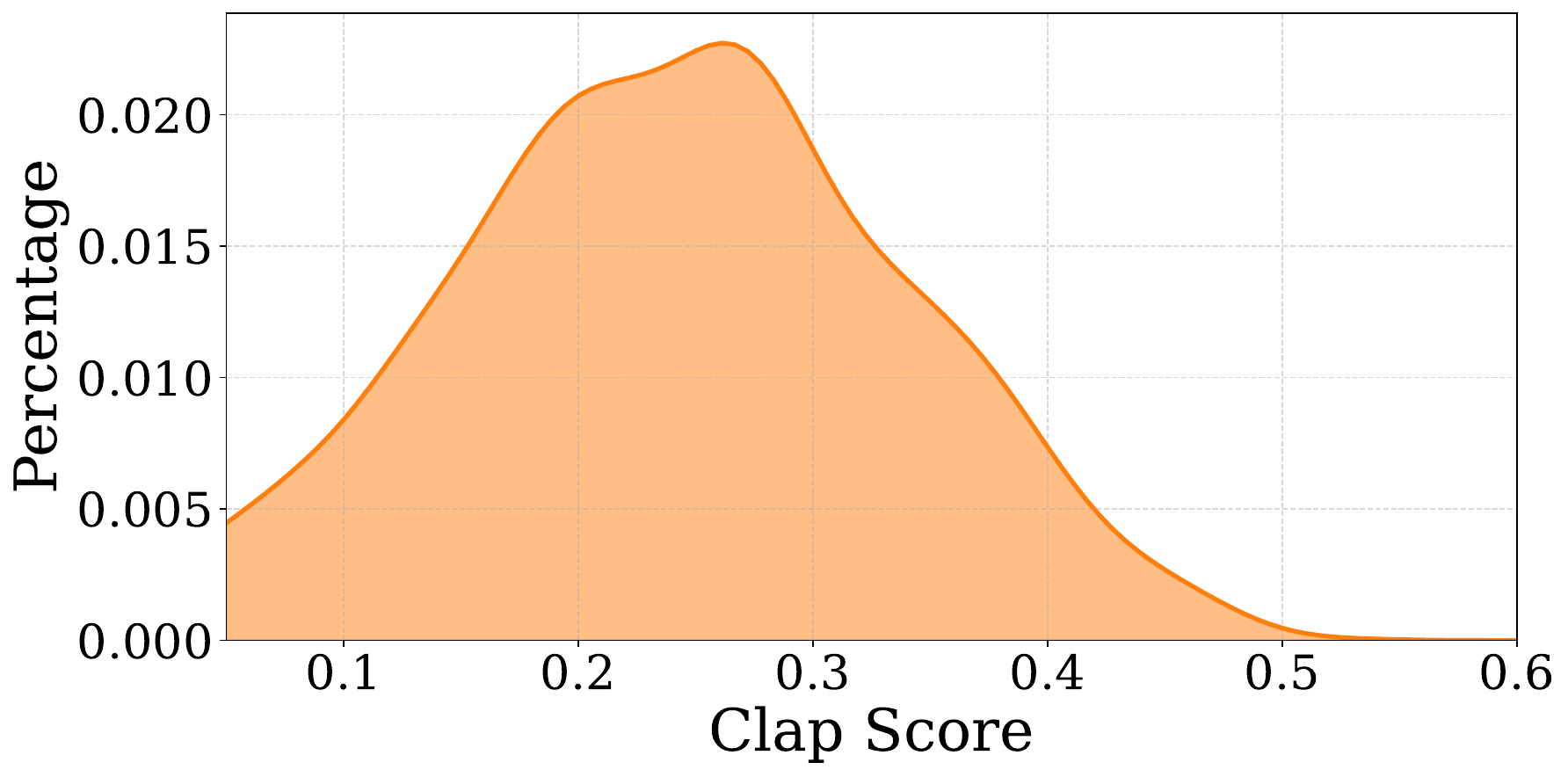}
                \caption{CLAP Score Distribution} 
                \label{fig:clap_score}
            \end{subfigure}
            \begin{subfigure}[b]{0.49\linewidth}
                \includegraphics[width=\linewidth]{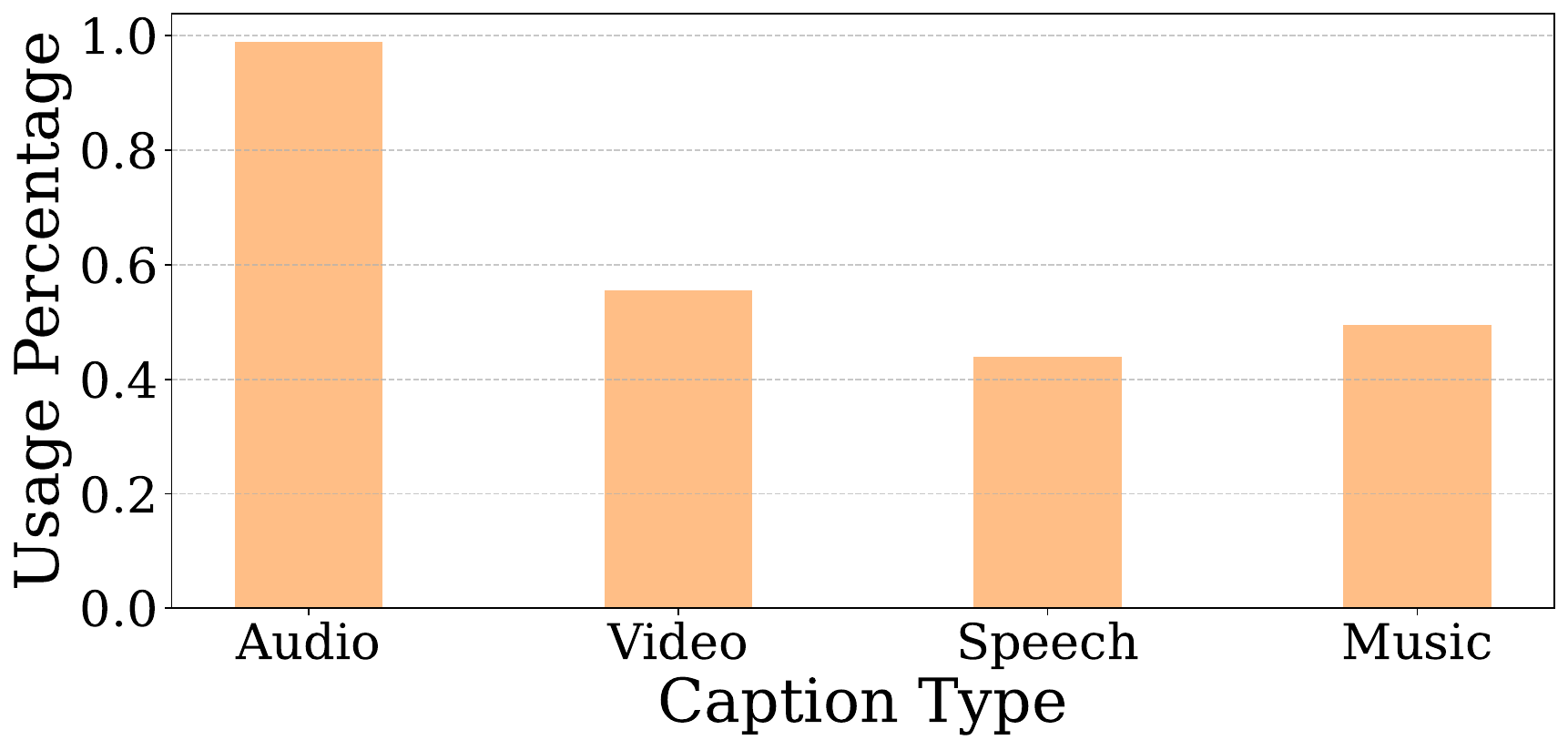}
                \caption{Modal Usage Frequency} 
                \label{fig:modal_use}
            \end{subfigure}%

        \end{subfigure}
    }
    
    \caption{Key statistics of FusionAudio-1.2M: (a) Proportion of top 5 audio labels from AudioSet; (b) Caption length comparison with existing datasets; (c) Diversity of semantic content types; (d) Proportion of captions utilizing different modalities; (e) Distribution of audio-text similarity measured by CLAP.} 
    \label{fig:caption_analysis_combined} 
\end{figure}

\paragraph{Dataset Statistics}
We analyze FusionAudio-1.2M across several dimensions:
\begin{itemize}
    \item \textbf{Audio Category Distribution:} Figure~\ref{fig:pie} shows the top five audio categories from AudioSet~\cite{7952261}, with "Music" being the most prevalent.  
    \item \textbf{Caption Length:} Figure~\ref{fig:caption_length} compares caption lengths (in tokens) with AudioCaps~\cite{kim-etal-2019-audiocaps}, Sound-VECaps~\cite{yuan2025soundvecapsimprovingaudiogeneration} and Auto-ACD~\cite{sun2024autoacdlargescaledatasetaudiolanguage}. FusionAudio-1.2M captions are significantly longer, indicating greater descriptive richness.
    \item \textbf{Semantic Diversity:} To showcase and compare the richness of semantic information across different datasets, we identified the presence of \textit{instruments}, \textit{emotions}, and \textit{music genres} in each caption using GPT-4o-mini (prompts in Appendix~\ref{app:Object Extraction Prompt}). Figure~\ref{fig:obj_type} shows FusionAudio-1.2M has higher coverage across most categories.
    \item \textbf{Audio-Text Alignment:} Figure~\ref{fig:clap_score} shows the distribution of cosine similarity between audio and text embeddings calculated by CLAP~\cite{laionclap2023}. Samples of different similarity scores can be found in Appendix~\ref{app:clap_score_samples}.
    \item \textbf{Modality Usage:} To better understand the contribution of different modalities to the final caption content, we use GPT-4o-mini to automatically annotate each caption for explicit references to different modalities: \textit{audio events}, \textit{speech}, \textit{music}, and \textit{visual context} (see prompt in Appendix~\ref{app:Object Extraction Prompt}). As shown in Figure~\ref{fig:modal_use}, over 50\% of samples integrate information from two or more modalities, demonstrating effective multimodal fusion. 

\end{itemize}


\subsection{Qualitative Analysis  }


\begin{figure}[htbp] 
    \centering 
    \setlength{\tabcolsep}{0pt} 
    
    \begin{subfigure}[b]{0.23\textwidth} 
        \centering
        \includegraphics[width=\textwidth]{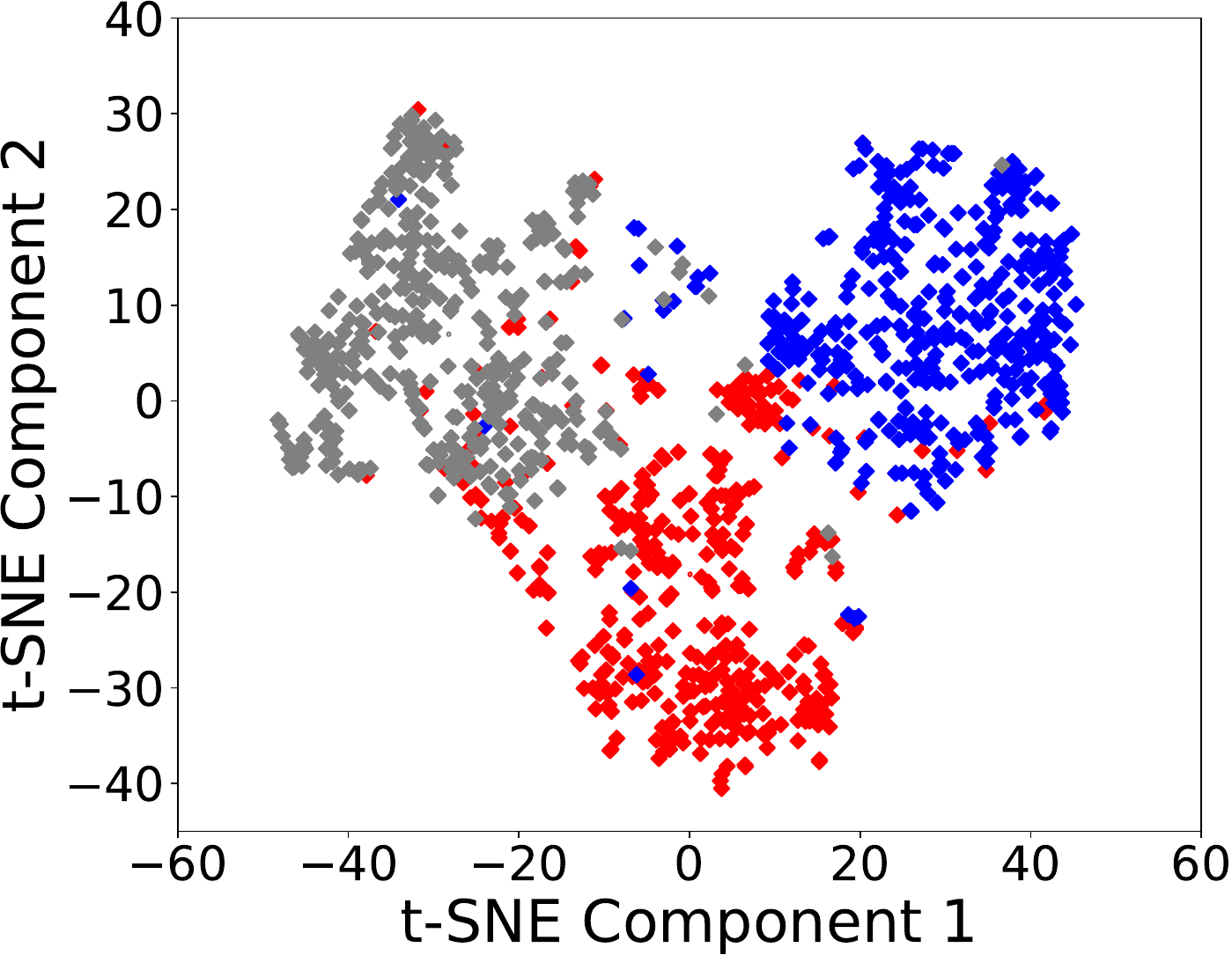}
        \caption{FusionAudio} 
        \label{fig:sub1}
    \end{subfigure}
    \hfill 
    \begin{subfigure}[b]{0.23\textwidth}
        \centering
        \includegraphics[width=\textwidth]{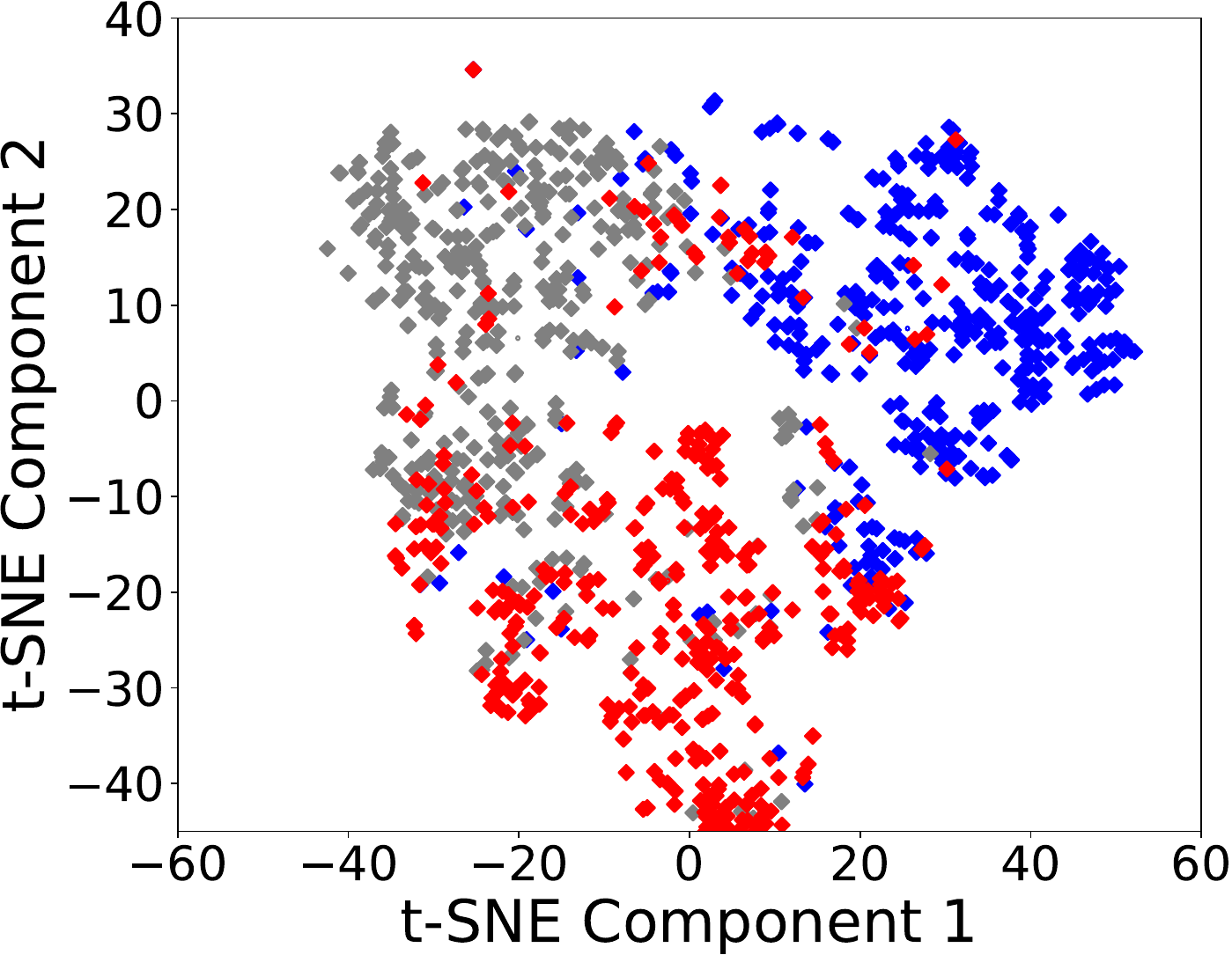}
        \caption{AudioSetCaps} 
        \label{fig:sub2}
    \end{subfigure}%
    \hfill
    \begin{subfigure}[b]{0.23\textwidth}
        \centering
        \includegraphics[width=\textwidth]{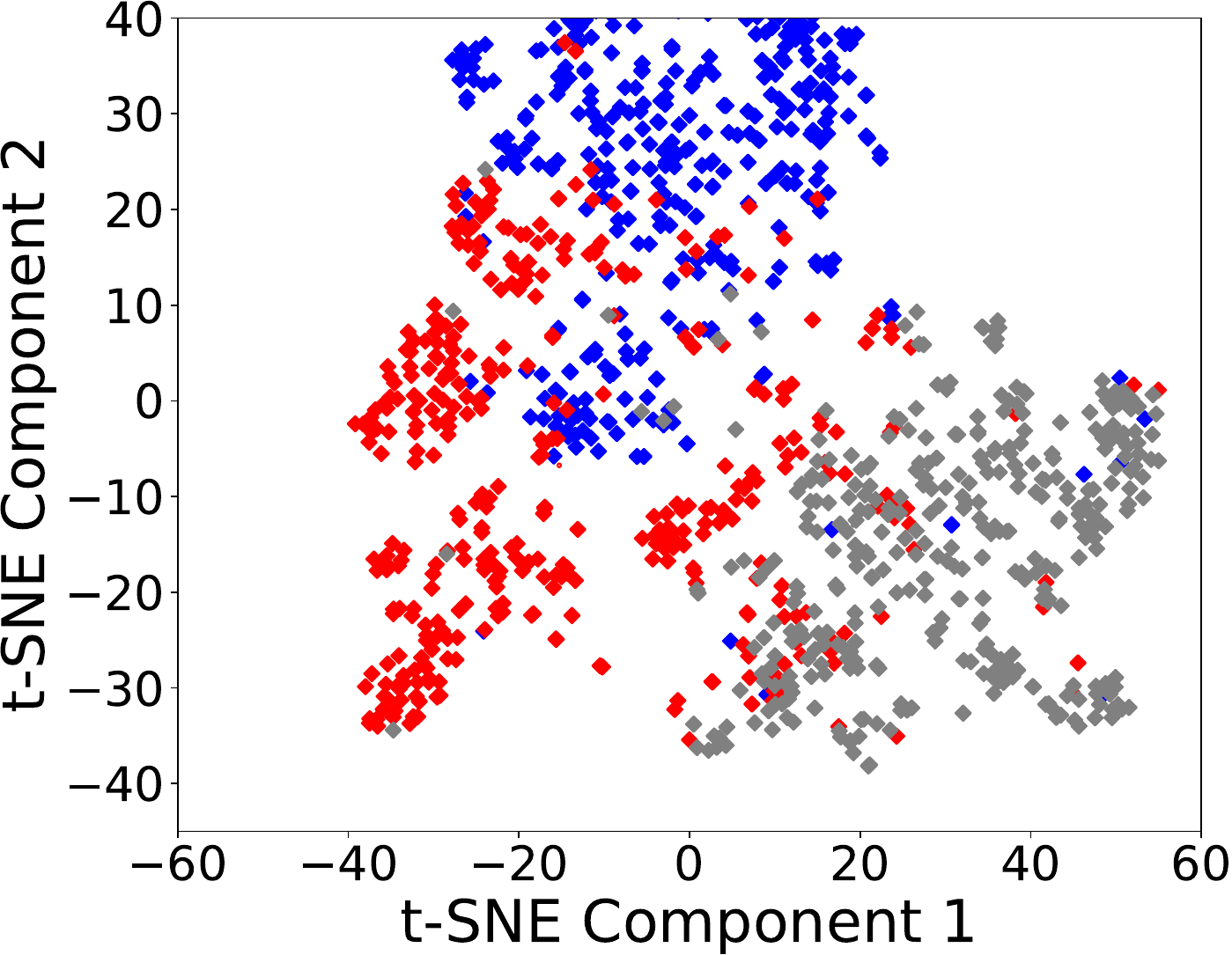}
        \caption{Auto-ACD} 
        \label{fig:sub3}
    \end{subfigure}%
    \hfill
    \begin{subfigure}[b]{0.23\textwidth}
        \centering
        \includegraphics[width=\textwidth]{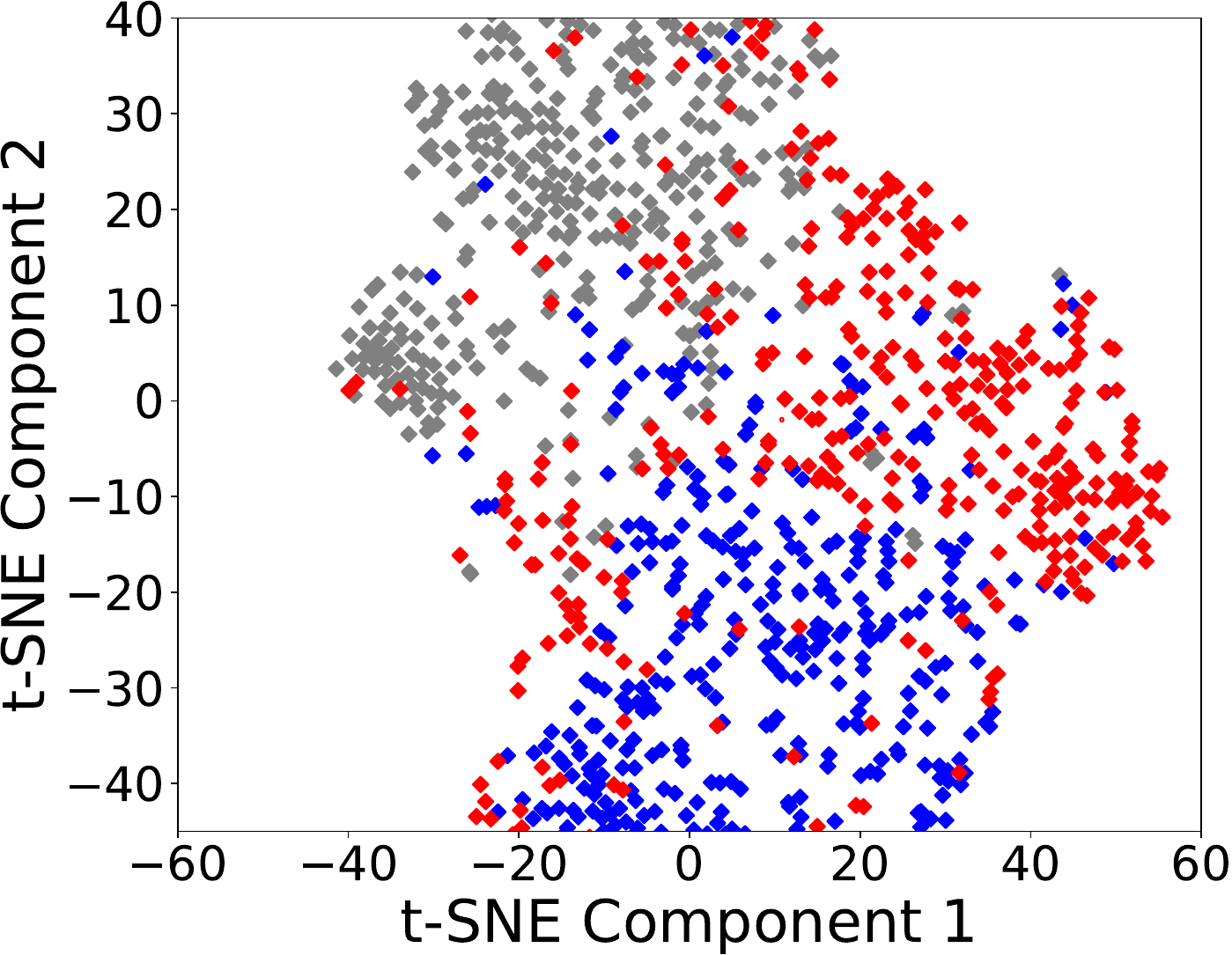}
        \caption{Sound-VECaps} 
        \label{fig:sub4}
    \end{subfigure}
    \vspace{0.001cm} 
    \centerline{ 
        \begin{subfigure}[b]{0.4\textwidth} 
            \centering
            \includegraphics[width=\textwidth]{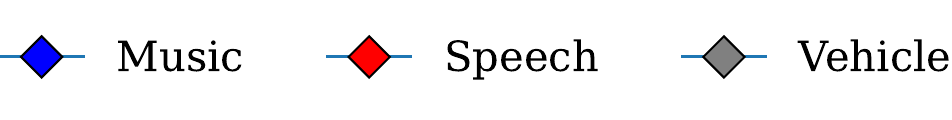} 
        \end{subfigure}
    }
    \captionsetup{skip=0pt}
    \caption{T-SNE Embedding of popular categories between different datasets} 
    \label{fig:embedding_tsne} 
    
\end{figure}

\textbf{Case Study}~
To further highlight the qualitative improvements enabled by FusionAudio, Table~\ref{tab:comparison} compares captions generated for the same audio clip across different datasets. FusionAudio's caption not only describes the primary sound event but also integrates visual cues, inferred context, and emotional tone, demonstrating a level of detail and reasoning absent from prior datasets. More samples can be found in Appendix~\ref{app:samples}.

\textbf{Embedding Projection for Visualizing Semantic Granularity}~
Embedding projection techniques like t-SNE~\cite{JMLR:v9:vandermaaten08a} visually reveal a dataset's semantic structure, illustrating intra-class compactness and inter-class separability—key indicators of data quality for discriminative tasks. We applied this to FusionAudio-1.2M by projecting CLAP sentence embeddings of its captions and those from baseline datasets into two dimensions using t-SNE. Figure~\ref{fig:embedding_tsne} demonstrates that FusionAudio's captions form significantly more compact same-category clusters and exhibit greater separation between different categories compared to baselines. This visual evidence indicates FusionAudio's superior semantic granularity and discriminative power, beneficial for nuanced audio understanding and cross-modal retrieval. Quantitative validation of inter- and intra-class distances is in Appendix~\ref{app:qutatitave_embedding}.

\section{Applications of FusionAudio-1.2M}
\label{sec:application}
We use FusionAudio-1.2M for two popular downstream tasks:  audio-text retrieval in Sec.~\ref{sec:retrieval} and audio understanding in Sec.~\ref{sec:understanding}. All experiments were conducted on a server equipped with 8 NVIDIA A800 80GB GPUs. Fine-tuning for the main experiments and ablation studies on the 25K data subset typically saved a checkpoint in under 30 minutes per run. The full evaluation process across all benchmark tasks required approximately 5 hours to complete per model evaluation. For the scaling study involving LLM fine-tuning, training sessions lasted between 10 to 12 hours.
\definecolor{highlightgray}{gray}{0.92}
\subsection{Audio-text Retrieval}
\label{sec:retrieval}

\begin{table}[htbp]
    \centering
    \caption{Audio-text retrieval performance (R@k, \%) on the AudioCaps test set.}
    \small
    \begin{tabular}{lc ccc ccc c}
        \toprule
        \textbf{\multirow{2}{*}{Dataset}} & \textbf{\multirow{2}{*}{Model}} & \multicolumn{3}{c}{\textbf{Text - to - Audio}} & \multicolumn{3}{c}{\textbf{Audio - to - Text}} & \textbf{\multirow{2}{*}{Avg.}} \\
        \cmidrule(lr){3 - 5} \cmidrule(lr){6 - 8}
        & & R@1 & R@5 & R@10 & R@1 & R@5& R@10 \\ 
        \midrule
        AC+CL & HTSAT+BERT & 36.1 & 71.8 & 83.9 & 46.8 & 82.9 & 90.7 & 68.7\\
        WavCaps & HTSAT+BERT & 42.2 & 76.5 & 87.1 & 54.6 & 85.2 & 92.4 & 73.0\\
        AudioSetCaps & HTSAT+BERT & 43.4 & 78.4 & 88.2 & 57.3 & 84.2 & 93.2 & 74.1\\
        Auto-ACD & HTSAT+RoBERTa & 42.7 & - & 88.5 & 56.3 & - & 93.9 & -\\
        Sound-VECaps & HTSAT+RoBERTa & 39.2 & 74.1 & 85.0 & 54.0 & 82.5 & 93.2  & 71.3\\
        \addlinespace
        \rowcolor{highlightgray} 
        FA(Ours) & HTSAT + BERT & \textbf{44.3} &  \textbf{79.9} &  \textbf{90.4} &  \textbf{57.8} &  \textbf{86.1} & \textbf{94.4}  & \textbf{75.5}\\
        \bottomrule
    \end{tabular}
    \label{tab:retrieval_result}
\end{table}

\subsubsection{Experimental Setup}
\paragraph{Tasks and Models}
We evaluate the quality of FusionAudio-1.2M by assessing its effectiveness as a pre-training corpus for the downstream task of cross-modal audio-text retrieval. This task requires retrieving the most relevant audio clip for a given textual query (text-to-audio retrieval) and, conversely, identifying the most pertinent text description for a given audio input (audio-to-text retrieval). For all experiments, we employ the HTSAT~\cite{chen2022hts}-BERT~\cite{devlin2019bert} model architecture.

\paragraph{Two-Stage Training}
Our training methodology for all evaluated datasets, including FusionAudio-1.2M and the baselines, follows a consistent two-stage protocol:
\begin{itemize}
    \item \textbf{Pre-training:} The HTSAT-BERT model is first pre-trained on the entirety of the respective source dataset (e.g., FusionAudio-1.2M, WavCaps, etc.). This stage utilizes a contrastive learning objective. For pre-training, the learning rate is set to 5e-5, the batch size is 196,training proceeds for 15 epochs.
    \item \textbf{Fine-tuning:} Subsequently, the pre-trained model undergoes full-parameter fine-tuning on the official training split of the AudioCaps (AC) dataset~\cite{kim-etal-2019-audiocaps}. For this fine-tuning stage, we use a learning rate of 1e-5, a batch size of 196, and train for 20 epochs.
\end{itemize}

\paragraph{Evaluation Setting}
The performance of all models, after the two-stage training protocol, is evaluated on the official test set of the AudioCaps dataset~\cite{kim-etal-2019-audiocaps}. We report Recall@k (R@k) for k=\{1, 5, 10\} for both text-to-audio and audio-to-text retrieval directions. R@k quantifies the percentage of queries for which the ground-truth item is successfully retrieved within the top-k ranked results. The comparative results are presented in Table~\ref{tab:retrieval_result}.

\subsubsection{Performance Analysis}
The detailed comparison results of model evaluation are presented in Table \ref{tab:retrieval_result}. The models trained on our dataset have significant advantages in the recall metrics of audio and text,which achieves the highest score in each R@k among models trained by existing audio caption datasets.The excellent audio and text recall performance indicates that the audio captions of FusionAudio-1.2M can accurately capture the information in the audio, ensuring the model's ability to distinguish fine-grained information. As a result, it can achieve high-accuracy matching even when dealing with similar audio.

\subsection{Audio Understanding}
\label{sec:understanding}
\definecolor{highlightgray}{gray}{0.92}

To empirically validate the practical utility and superior quality of our proposed FusionAudio-1.2M dataset, we evaluated its impact on a comprehensive suite of audio understanding tasks. Specifically, we benchmarked the performance of the GAMA model~\cite{ghosh-etal-2024-gama} fine-tuned on FusionAudio-1.2M against instances of the same model fine-tuned on several established audio captioning datasets.

\begin{table}[htbp]
\centering
\caption{Performance comparison of the GAMA model fine-tuned on FusionAudio-1.2M against baseline datasets across a battery of audio understanding evaluation benchmarks. All models were fine-tuned on 25,000 QA pairs. M.J. denotes Model-Judge score (using GPT-4.1-mini). Best scores are in bold. The three main categories of evaluation tasks align with those in Table~\ref{tab:captioning_scenarios}.}
\scriptsize
\setlength{\tabcolsep}{2pt}
\captionsetup{font=scriptsize}

\begin{tabular}{l|ccccc|cccccccc|ccccc}
\toprule
\textbf{\multirow{3}{*}{\scriptsize Dataset}} &
    \multicolumn{5}{c|}{\scriptsize \textbf{Adverse Acoustic Conditions}} & \multicolumn{8}{c|}{\scriptsize \textbf{High-Level Semantic Understanding}} & \multicolumn{5}{c}{\scriptsize \textbf{Fine-grained Information}} \\

 & \makecell[c]{AS\\{\tiny (Acc.)}} & \makecell[c]{US$_{\text{8k}}$\\{\tiny (mAP)}} & \makecell[c]{TAU\\{\tiny (mAP)}} & \makecell[c]{FSD$_{\text{ns}}$ \\{\tiny (mAP)}} & \makecell[c]{Avg.}  & \makecell[c]{Genre\\{\tiny (Acc.)}} & \makecell[c]{M$_{\text{AQA}}$\\{\tiny (Acc.)}} & \makecell[c]{Mood\\{\tiny (Acc.)}} & \makecell[c]{M$_{\text{chat}}$\\{\tiny (M.J)}}& \makecell[c]{S$_{\text{AQA}}$\\{\tiny (Acc.)}}  & \makecell[c]{S$_{\text{chat}}$\\{\tiny (M.J)}}& \makecell[c]{AB$_{\text{Sc}}$\\{\tiny (M.J)}} & \makecell[c]{Avg.} & \makecell[c]{Vocal\\{\tiny (Acc.)}} & \makecell[c]{Instr\\{\tiny (Acc.)}} & \makecell[c]{ESC\\{\tiny (Acc.)}} & \makecell[c]{FSD\\{\tiny (mAP)}} & \makecell[c]{Avg.} \\

\midrule

GAMA(base)         & 48.0 & 56.6        & 23.5 & 81.9 & 52.5  & 42.8 & 44.1 & 28.3 & 45.4& 50.1 & 58.9 & 56.0 & 46.5 & 63.5 & 68.7 & 68.9 & 45.8 & 61.7 \\
AC+CL              & 50.3 &\textbf{65.3}& 21.3 & 81.9 & 54.7  & 49.4 & 50.7 & 28.4 & 47.0& 52.3 & 55.4 & 61.3 & 49.2 & 68.2 & 68.9 & 65.7 & 39.9 & 60.8 \\
WavCaps            & 55.4 & 64.5        & 25.0 & 77.6 & 55.6  & 53.4 & 51.6 & 33.2 & 27.7& 45.1 & 29.7 & 52.7 & 41.9 & 55.4 & 69.7 & 58.8 & 32.4 & 54.1\\
ASC                & 45.4 & 51.3        & 22.3 & 77.8 & 49.2  & 56.0 & 57.6 & 31.6 & 51.3& 49.9 & 59.5 & 58.5 & 52.1 & 51.8 & 70.5 & 57.7 & 30.5 & 52.6 \\
CompA-R            & 56.5 & 63.3        & 22.7 & 83.7 & 56.6  & 60.1 & 54.7 & 33.9 & 47.0& 56.1 & 58.3 & 60.1 & 52.9 & 63.5 & 68.6 & 62.3 & 38.4 & 58.2 \\
\addlinespace
\rowcolor{highlightgray} 
\textbf{FA(ours)}  & 59.0 & 58.8 & 24.4   & 84.6  & 56.7  & \textbf{65.1} & 57.6 & 35.7 & 57.1& \textbf{59.1} & 61.5 &\textbf{64.5} & 57.4 & 69.0 & 73.6 & 65.5 & 44.5 & 63.0 \\
\rowcolor{highlightgray} 
\textbf{FA-high(ours)} & \textbf{59.7} &64.0& \textbf{25.1} & \textbf{88.2} & \textbf{59.3}  & 64.2 & \textbf{60.0} & \textbf{38.3} & \textbf{57.9}& 58.4 & \textbf{62.3} & 64.0 & \textbf{57.9} & \textbf{71.0} & \textbf{73.9} & \textbf{71.3} & \textbf{47.4} & \textbf{65.9} \\
\bottomrule
\end{tabular}

\captionsetup{font=normalsize} 
\label{tab:audio_benchmark}
\end{table}

\subsubsection{Experimental Design}
\label{ssec:exp_design}

\paragraph{Tasks and Models} We focused on general audio understanding beyond speech, employing the GAMA model architecture~\cite{ghosh-etal-2024-gama}, a transformer-based audio-language model, as our foundation for fine-tuning. Fine-tuning utilized a learning rate of 5e-5, a batch size of 128, and 2 training epochs. Evaluation utilized t=0.1 for inference.
 
\paragraph{Training} The GAMA model was fine-tuned independently on several datasets: our FusionAudio-1.2M and its high-quality subset FusionAudio-high (top 25k QA pairs selected for quality and diversity), alongside established datasets. A critical aspect was \textbf{normalizing training data to 25,000 QA pairs} across all datasets, ensuring performance differences primarily reflect data quality, not quantity. Notably, while baseline datasets typically required 25,000 unique audio clips (one QA pair per clip) for this volume, FusionAudio-1.2M achieves this with only \textbf{9,000 unique audio clips}, owing to its design of multiple rich QA pairs per audio instance.

\paragraph{Evaluation} Fine-tuned models were evaluated on 15 diverse audio understanding tasks (Table~\ref{tab:audio_benchmark}), assessing capabilities across three key scenarios: (1) robustness to Adverse Acoustic Conditions, (2) proficiency in High-Level Semantic Understanding, and (3) acuity in discerning Fine-grained Information. For benchmarks requiring automated judgment (M.J. scores in Table~\ref{tab:audio_benchmark}), we employed \textit{GPT-4.1-mini}.

\subsubsection{Performance Analysis}
\label{ssec:perf_highlights}

The results in Table~\ref{tab:audio_benchmark} demonstrate the significant advantages of fine-tuning with FusionAudio. 

\paragraph{Dominant Performance Driven by High-Quality and Efficient Data}
GAMA fine-tuned on FusionAudio, and especially its FusionAudio-high subset, consistently outperformed models trained on all benchmarked datasets across the majority of the 13 tasks, with FusionAudio-high achieving the highest average scores in all scenarios. This success is rooted in the superior intrinsic quality of FusionAudio, crafted by our method to maximize information richness per audio clip. As a direct result, models can learn more effectively from each sample, leading to the crucial observation that this dominant performance was achieved using substantially fewer unique audio clips than other datasets. This clearly demonstrates that FusionAudio-1.2M not only provides higher-caliber data overall but also facilitates more efficient learning and utilization of each individual audio piece.

\section{Ablation Study}
\subsection{On the Effectiveness of Multimodal Cues}

To rigorously evaluate the contribution of each component in our method for enhanced audio information, we conducted a comprehensive ablation study. This study aims to (1) ascertain the individual importance of each auxiliary modality (Speech, Music, Video) in augmenting the Sound modality, and (2) validate the effectiveness of our proposed automatic filtering module.

\paragraph{Experiment Setup} All ablation experiments were performed on the same subset from AudioSet with a scale of 25k, using the same audio clips and training procedures. FusionAudio-1.2M incorporates all four modalities (Sound, Music, Speech, Video) and includes the multi-modal fusion quality threshold filtering module. We compared FusionAudio-1.2M against several ablated variants.

\paragraph{Ablation Results on Fusion} As shown in Table \ref{tab:ablation_study}, ablating auxiliary modalities (Music, Video, Speech) generally degraded performance. Removing video captions (w/o Video) caused the most significant decline, underscoring visual context's critical role. Ablating music (w/o Music) and speech (w/o Speech) also reduced performance. An interesting exception was observed for Task 1, where removing speech (w/o Speech) led to a slight improvement. We attribute this to a combination of potentially poor ASR transcription quality in adverse acoustic conditions, which could introduce detrimental noise, and a possible task focus shift where non-speech acoustic analysis is prioritized, making speech content less critical and potentially diverting optimization from core modalities. Notably, the magnitude of these performance drops (-0.76 for Music, -1.18 for Video, and -0.93 for Speech on average) generally corresponds with the usage of these modalities in our dataset, as illustrated in Figure \ref{fig:modal_use}. This suggests that modalities more frequently leveraged for information contribute more significantly to the overall performance.

\paragraph{Ablation Results on Filtering} Finally, removing our quality filtering module led to a consistent, significant performance drop across all tasks, highlighting its effectiveness in mitigating issues from hallucinations introduced during the process.

\subsection{On the Effectiveness of Data Scaling} 
To assess the impact of data volume, we conducted scaling experiments for the downstream tasks. This study evaluates performance gains as data size increases, providing insights into model scalability.

\paragraph{Experiment Setup} We used nested subsets, starting from 1.25K audio clips. The Audio Understanding task scaled to 80k clips (355k QA pairs), while Retrieval utilized up to the full 1.2M clips. Model architectures and training hyperparameters remained consistent with previous experiments.


\begin{table}[htbp!]
\centering
\caption{Ablation Study on FusionAudio-25K Dataset. Performance metrics are shown for Retrieval tasks (Text-to-Audio and Audio-to-Text) and Understanding tasks (\textit{Task I}: Adverse Acoustic Conditions; \textit{Task II}: High-level Semantic Understanding; \textit{Task III}: Fine-grained Information).}
\small
\label{tab:ablation_study} 
\begin{tabular}{l| cc| ccc|c}
\toprule
  \multirow{2}{*}{\small \textbf{Settings}} &
  \multicolumn{2}{c|}{\small \textbf{Retrieval Task}} & 
  \multicolumn{3}{c|}{\small \textbf{Understanding Task}} &
  \multirow{2}{*}{\small \textbf{Avg.}}\\
  &
  \makecell[c]{\small T-A} & 
  \makecell[c]{\small A-T} & 
  \makecell[c]{\small \textit{Task I}: AAC} & 
  \makecell[c]{\small \textit{Task II}: HSU} &
  \makecell[c]{\small \textit{Task III}: FI} &
  \\
\midrule
\rowcolor{highlightgray}
FusionAudio-1.2M            & \textbf{39.70} & \textbf{49.71} & 56.73 & \textbf{57.16} & \textbf{63.02} & \textbf{53.26} \\
\quad w/o Music        & 39.03 & 47.53 & 56.72 & 56.34 & 62.87 & 52.50(-0.76) \\
\quad w/o Video        & 38.53 & 48.79 & 55.90 & 56.12 & 61.08 & 52.08(-1.18) \\
\quad w/o Speech       & 38.09 & 47.87 & \textbf{57.38} & 56.06 & 62.27 & 52.33(-0.93) \\
\quad w/o Filter       & 39.45 & 49.14 & 55.30 & 55.25 & 61.35 & 52.10(-1.16) \\
\bottomrule
\end{tabular}
\end{table}

\begin{figure}[htbp!] 
    \centering 
    \setlength{\tabcolsep}{0pt} 
    
    \begin{subfigure}[b]{0.48\textwidth} 
        \centering
        \includegraphics[width=\textwidth]{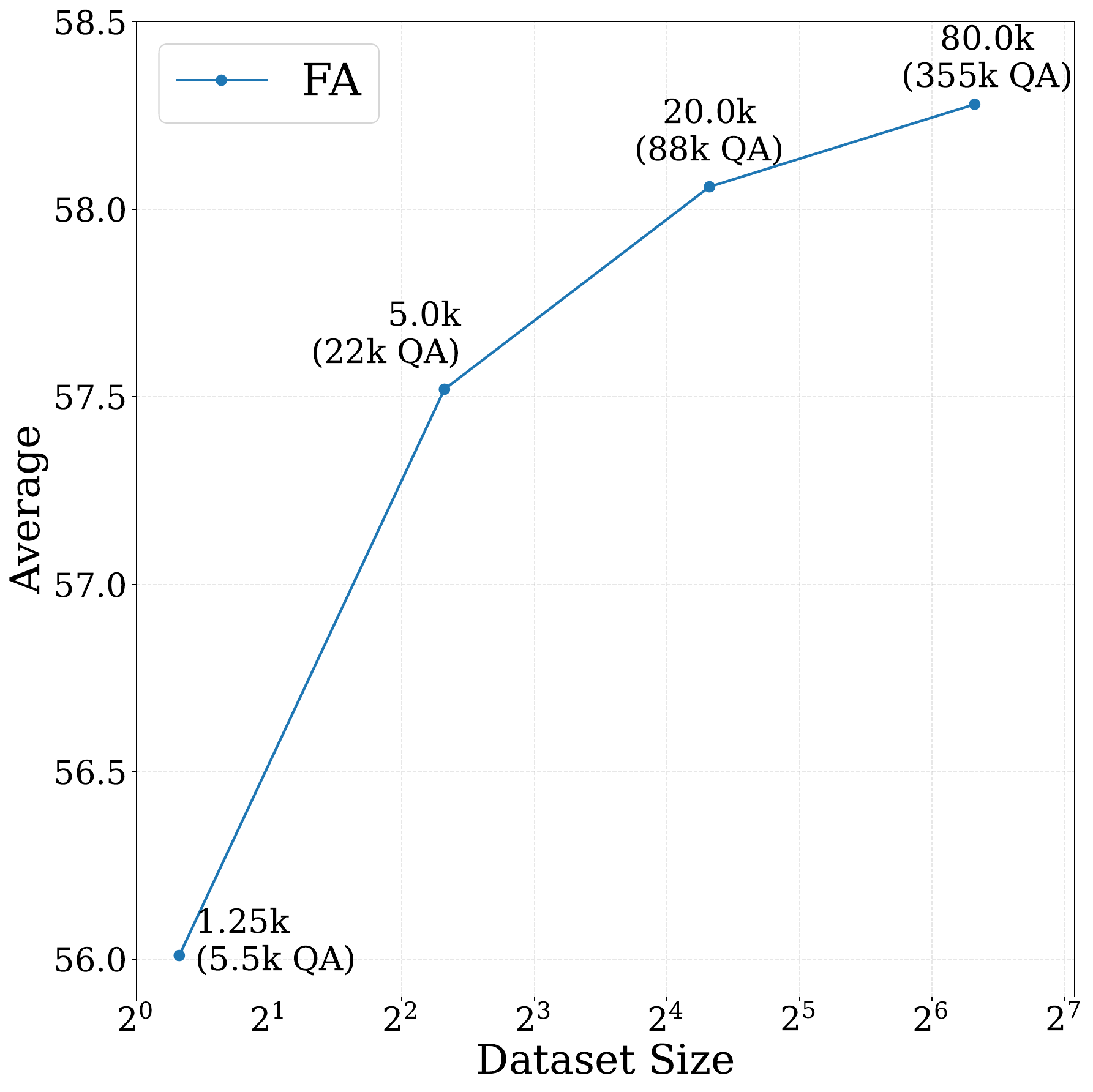}
        \caption{Audio Understanding} 
        \label{fig:understanding_scaling.pdf}
    \end{subfigure}
    \hfill 
    \begin{subfigure}[b]{0.48\textwidth}
        \centering
        \includegraphics[width=\textwidth]{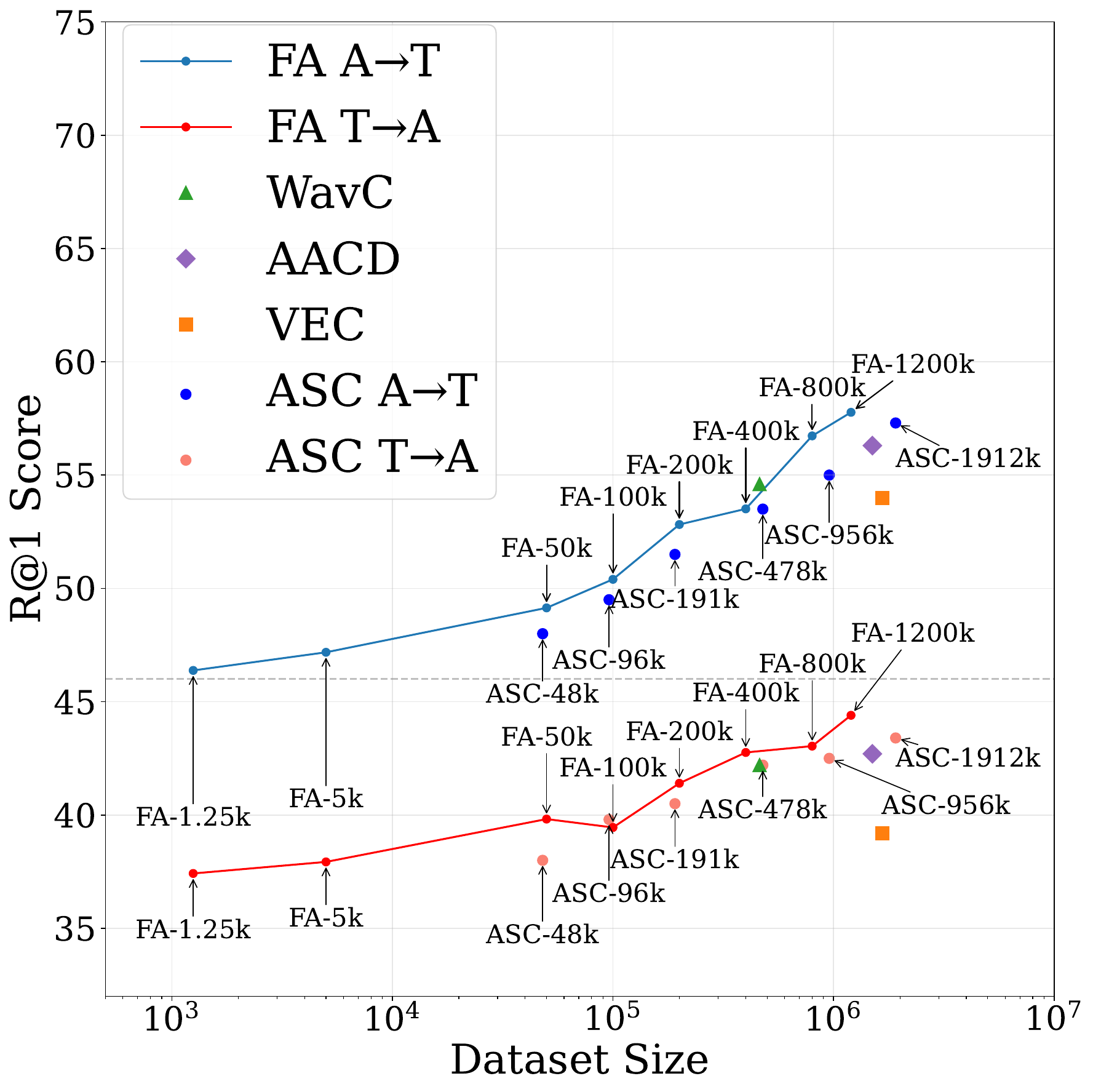}
        \caption{Audio-text Retrieval} 
        \label{fig:audio_text_sub2}
    \end{subfigure}
    \captionsetup{skip=8pt} 
    \caption{Scaling result of understanding and retrieval tasks. Details of the legend in (b): \\A: Audio; T: Text; FA: FusionAudio-1.2M; WavC: WavCaps; AACD: Auto-ACD; VEC: Sound-VECaps; ASC: AudioSetCaps.} 
    \label{fig:scaling_result} 
\end{figure}

\paragraph{Results} As depicted in Figure \ref{fig:scaling_result}, increasing data volume consistently improved performance for both tasks. For Audio Understanding, scaling from 1.25K to 80k clips enhanced average performance, likely due to increased exposure to diverse audio and associated QA pairs generated with a natural distribution. The Retrieval task exhibited a substantial and consistent rise in Recall@1 with data expansion, peaking with the full dataset. These findings underscore that greater data volume generally boosts model capabilities, highlighting the value of our dataset's scale and richness.

\section{Conclusion}

This paper presents FusionAudio-1.2M, a large-scale dataset for fine-grained audio captioning created via a novel multimodal contextual fusion pipeline. Inspired by human auditory perception, the approach combines specialized expert models for speech, music, sound events, and visual context with LLM-based synthesis. 
Experiments show that models trained on FusionAudio-1.2M achieve strong performance using fewer unique audio samples due to richer per-clip annotations. Ablation studies confirm the significance of each modality, particularly visual context. 
This work could be further improved by  polishing the caption generation method, diversifying the dataset with longer audio clips, probing more sophisticated multimodal fusion techniques, and performing a deeper societal impact analysis, which we leave as future work.

\section*{Ackownledgement}
This work was supported by the Shenzhen Science and Technology Program (JCYJ20220818103001002), Shenzhen Doctoral Startup Funding (RCBS20221008093330065), Tianyuan Fund for Mathematics of National Natural Science Foundation of China (NSFC) (12326608), Shenzhen Science and Technology Program (Shenzhen Key Laboratory Grant No. ZDSYS20230626091302006), and Shenzhen Stability Science Program 2023, Shenzhen Key Lab of Multi-Modal Cognitive Computing.

\section*{Limitation}
\label{sec:limitation}
The study also acknowledges several limitations. First, the automated generation of audio captions may introduce hallucinations or errors, despite quality assurance measures such as human evaluation and automatic filtering. Second, the dataset primarily focuses on short audio clips (10 seconds), which may limit its applicability to longer or more complex audio contexts. Third, the multimodal fusion approach relies on the integration of speech, music, visual, and general audio information, but the interplay and weighting of different modalities are not thoroughly explored. Fourth, due to computational resource constraints, we were unable to conduct multiple experimental runs to establish robust error bars for all reported metrics, which could provide further statistical confidence. Lastly, while our pipeline incorporates a quality filtering module to mitigate LLM-induced inaccuracies, completely eliminating potential hallucinations in automated data generation remains an ongoing challenge, suggesting a need for continued refinement in robust AI-driven data creation.\\
Future work could address these limitations by further refining the caption generation process, expanding the dataset to include longer audio clips, exploring more nuanced multimodal fusion strategies, and conducting a more comprehensive analysis of societal impacts.

\newpage
\bibliography{ref}
\bibliographystyle{unsrt}

\newpage
\appendix

\section{Human Evaluation}
\label{app:human_eval}
\subsection{Evaluation Setup}
We recruit five evaluators to assess the data. All evaluators are students with a bachelor's degree or higher and have studied in an English-only teaching environment. The five evaluators are tasked with evaluating a total of 300 samples. Each evaluator is assigned 120 samples, ensuring that each sample is evaluated twice by different evaluators. 

Evaluators are required to score the captions based on two dimensions: the level of detailness and the degree of hallucination. 

\begin{itemize}
    \item \textbf{Detailness:} Evaluating the level of detail, specificity, and contextual information provided in the caption regarding the audio events and scene. Captions describing multiple relevant aspects accurately scored higher. Detailness is scored through 1-3.
    \item \textbf{Hallucination:} Assessing the accuracy of the description against the source audio-visual content. This specifically penalizes hallucinated objects, events, or attributes not perceivable in the clip. Hallucination is scored through 1-5.
\end{itemize}

Specific scoring guidelines can be found in the Appendix~\ref{app:human_eval_instructions}.

\subsection{Instruction for Human Evaluation}
\label{app:human_eval_instructions}
The instruction used for human evaluation is shown in Figure~\ref{fig:rating_instruction}.

\subsection{F-Score Computation}
\label{app:f_score}

To balance precision and recall in our automatic filtering process, we used the $F\textsubscript{1.05}$ score, which slightly emphasizes recall over precision. This emphasis ensures that captions with high hallucination rates are effectively discarded, even at the cost of filtering out some acceptable ones. The $F\textsubscript{1.05}$ score is calculated using the formula:

\[
F\textsubscript{1.05} = \frac{(1 + 1.05^2) \cdot \text{Precision} \cdot \text{Recall}}{(1.05^2 \cdot \text{Precision}) + \text{Recall}}
\]

Where precision and recall are computed from the confusion matrix as:

\[
\text{Precision} = \frac{\text{TP}}{\text{TP} + \text{FP}} \quad \text{and} \quad \text{Recall} = \frac{\text{TP}}{\text{TP} + \text{FN}}
\]

\begin{figure*}[!ht] 
\small
\begin{AIbox}{Instruction for Human Evaluation}

\section*{Introduction}
You are tasked with evaluating captions generated for audio clips. Please use the following guidelines to assess each caption based on two indicators: \textbf{Detailing} and \textbf{Hallucinations}

\section*{1. Detailing}

\section*{Key Things to Look For:}
\begin{itemize}
    \item Whether the caption captures all major sounds and events in the audio (e.g., dog barking, doorbell ringing, etc.).
    \item If the intensity or emotional context of the sound is conveyed (e.g., the dog barking intensely or the doorbell ringing in a rapid succession).
    \item Whether the caption includes additional information when relevant (e.g., a dog barking \textit{repeatedly} or \textit{distressed}).
\end{itemize}

\section*{Scoring Guidelines:}
Categorize captions into three detail levels (high, medium, low)based on their coverage of audio elements.
\begin{itemize}
    \item \textbf{Low:} Only generic descriptions without specific elements
    \item \textbf{Medium:} Identifies main elements but lacks contextual details
    \item \textbf{High:} Specifies sound sources, qualities, and relationships
\end{itemize}

\section*{2. Hallucinations}

You will be given the highlighted words or phrases marked by DeepSeek-V3 that need to be verified in the original caption:\\
2

\noindent A [\textbf{male voice}] delivers a [\textbf{scripted narration}] [\textbf{in Polish}], likely from a [\textbf{recorded radio or podcast segment}], accompanied by [\textbf{subtle studio ambiance}] including [\textbf{microphone hiss}] and [\textbf{paper rustling}]. A [\textbf{secondary listener}] [\textbf{wearing headphones}] remains [\textbf{audibly inactive}], though [\textbf{faint page-turning sounds}] indicate [\textbf{preparatory material review}]. The spoken text references [\textbf{program materials available at Lechia.net}], suggesting a [\textbf{structured broadcast format}] with [\textbf{editorial oversight}]. Background contains [\textbf{minimal environmental noise}] consistent with a [\textbf{sound-treated recording space}].  \\

\noindent Total flagged phrases: \textbf{17
}\\

\noindent \textbf{Note}: The total number of flagged phrases is provided for reference. If you believe other words or phrases are important in the context of the verification, please consider them in your calculation as well.

\subsection*{Your Task}

\begin{itemize}
    \item Listen to the audio and verify the highlighted elements.
    \item Assign one of the following error values to each phrase:
\end{itemize}

\[
\begin{array}{|l|l|}
\hline
\textbf{Label} & \textbf{Criteria} \\
\hline
\text{Correct (0)} & \text{Directly verifiable from audio} \\
\hline
\text{Unverifiable (0.5)} & \text{Neither confirmed nor disproven, or things you are not sure} \\
\hline
\text{Hallucination (1)} & \text{Contradicts audio or invents content} \\
\hline
\end{array}
\]

\subsection*{Scoring Calculation:}
The final hallucination rate is calculated as follows:

\[
\text{Hallucination Rate} = \left(\frac{\sum(\text{Error Values})}{\text{Total Content Units}}\right) \times 100\%
\]
Based on the hallucination rate, assign a final score as follows:

\textbf{0-10\%:} Score = 5 | 
\textbf{11-25\%:} Score = 4 | 
\textbf{25-40\%:} Score = 3 | 
\textbf{41-50\%:} Score = 2 | 
\textbf{51-100\%:} Score = 1 

\end{AIbox}
\caption{Instruction for Human Evaluation.}
\label{fig:rating_instruction}

\end{figure*}

\subsection{Human Rating Distribution}

\begin{figure*}[t]
    \centering 
    \begin{subfigure}[t]{0.9\textwidth} 
        \centering
        \includegraphics[width=\textwidth]{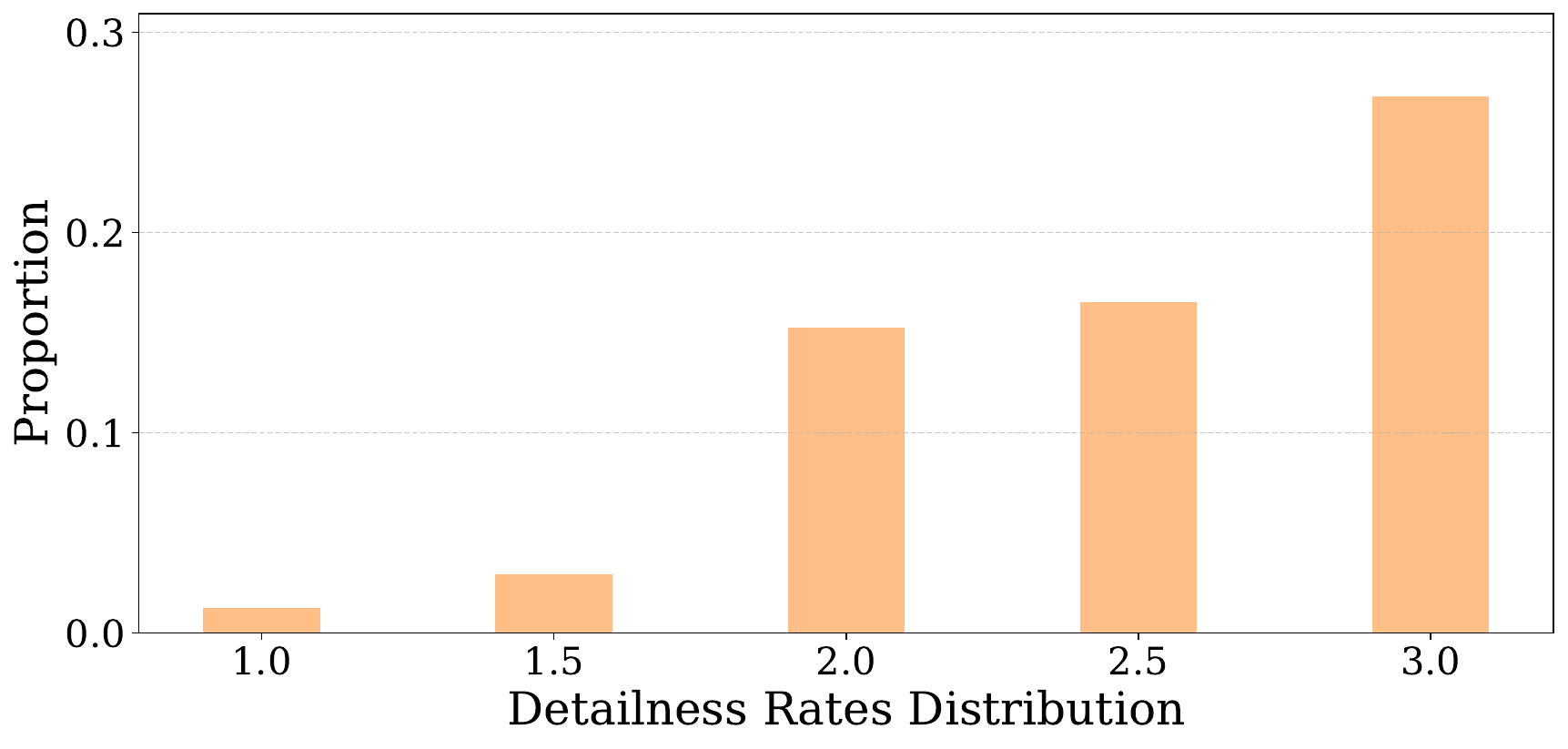}
        \label{fig:detailness}
    \end{subfigure}
    
    \vspace{-0.3cm} 
    
    \begin{subfigure}[t]{0.92\textwidth} 
        \centering
        \includegraphics[width=\textwidth]{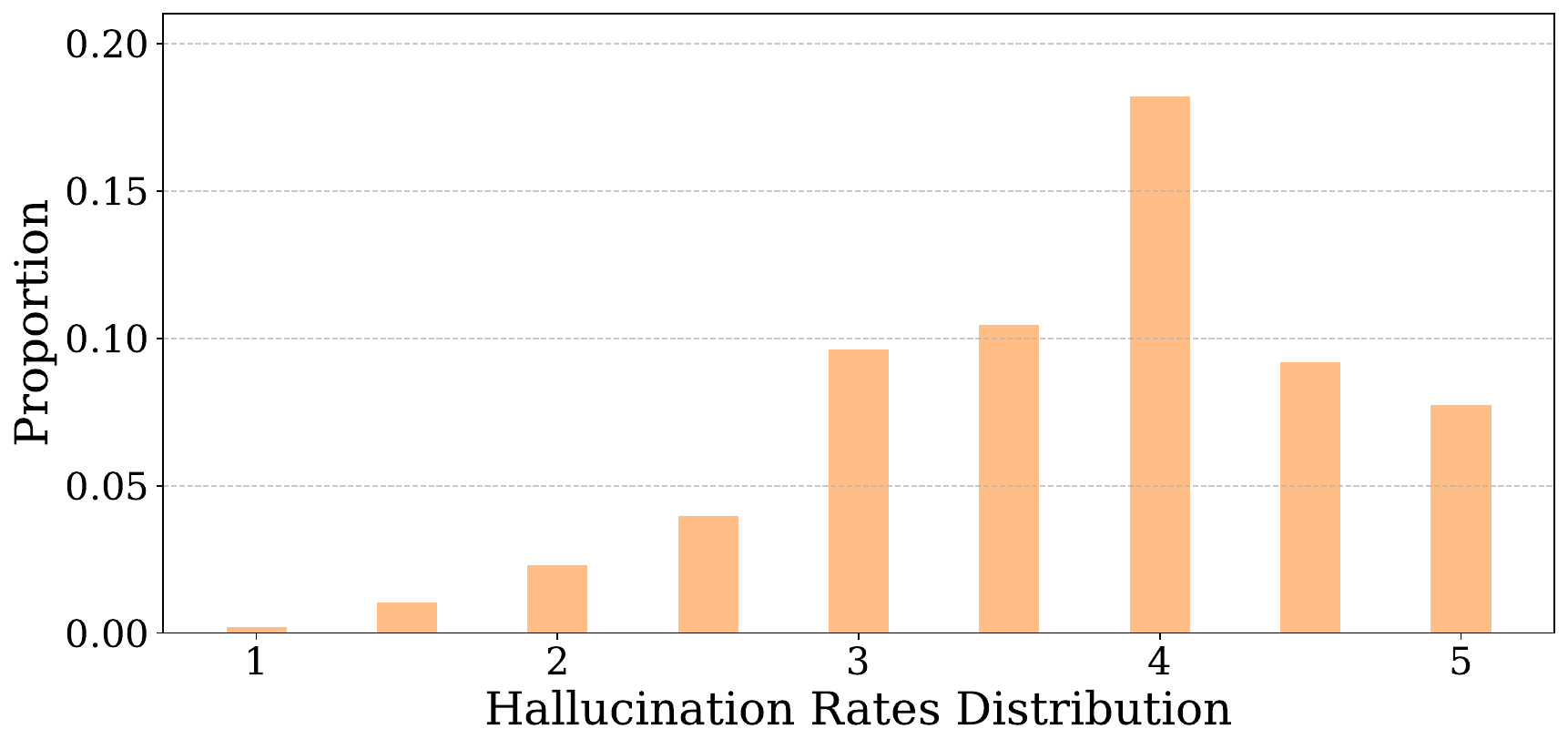}
        \label{fig:hallucination}
    \end{subfigure}
    \hspace{0.01\textwidth} 
    
    \vspace{-0.5cm}
    \setlength{\parindent}{10cm} 
    \caption{Detailness and Hallucination Rates Distribution of Human Rating}
    \label{fig:HumanRating}
\end{figure*}

We statistically analyze the distribution of human ratings for detailness and hallucination,which are shown as Figure~\ref{fig:HumanRating}.

\section{Prompt for models}
\label{app:prompts}
\subsection{Video Caption Prompt}
\label{app:visual_prompt}

\begin{figure*}[t]
\begin{AIbox}{Prompt for video caption}
\footnotesize
{\bf Prompt:} \\
Please provide a comprehensive video description focusing exclusively on observable visual elements, including timestamps:

**1. Key Entities \& Actions with Timestamps:** \\
- List main objects/subjects and their visible actions with approximate timestamps (MM:SS format)\\
- Describe: \\
  * Object/subject movements and interactions\\
  * Material properties (metal, wood, liquid)\\
  * Timing of significant visual events\\

**2. Scene Description with Timeline:**\\
- Overall scene dynamics and visual interactions\\
- Notable visual events with timestamps:\\
  * Object collisions or impacts\\
  * Movement patterns\\
  * Material changes\\
  * Human/animal visible actions\\
- Environmental context (indoor/outdoor, spatial relationships)\\

**3. Overall Description with Chronological Flow:**\\
- Provide a comprehensive visual narrative of the video\\
- Include timestamps for key moments and transitions (MM:SS format)\\
- Focus on observable actions, and movements\\
- Use specific, action-oriented language\\
- Present events in chronological order with clear time markers\\

Guidelines:\\
- Describe only directly visible elements\\
- Focus on observable actions and movements\\
- Note material properties and physical interactions\\
- Include **timestamps** for all significant events\\
- Timestamp **should not** exceed the duration of the video\\
- Use precise descriptive language for visual elements\\
- Avoid assumptions about non-visible elements\\
- Maintain strict focus on visual information\\

Example:\\
Instead of "A car's engine roars as it accelerates"\\
Write "00:01 - A red sports car with chrome detailing accelerates down a paved road, tires creating visible spray on wet asphalt"\\
"00:02 - The car's rear suspension compresses during acceleration, exhaust emitting visible vapor"\\
"00:03 - The car's engine roars as it accelerates"\\
\end{AIbox}
\caption{Prompt for video caption.}
\label{fig:visual_caption_prompt}

\end{figure*}

The prompt we used for Qwen2.5-VL-72B to extract video caption is shown in Figure~\ref{fig:visual_caption_prompt}. We try to let the model describe sound-related object only, but found that it would introduce additional hallucinations. Thus, we prompt the model to describe visual content only, and let the integration model tackle the modality issue.

\subsection{Audio Caption Prompt}
\label{app:audio_caption_prompt}

\begin{figure*}[t]
\begin{AIbox}{An example prompt for audio caption generation}
\footnotesize
Describe the audio in detail, but there is not need for association or speculation.
\end{AIbox}
\caption{An example prompt for audio caption generation}
\label{fig:audio_caption_prompt}

\end{figure*}

The prompt we used for GAMA to extract audio caption is shown in Figure~\ref{fig:audio_caption_prompt}.

\subsection{Object Extraction Prompt}
\label{app:Object Extraction Prompt}

\begin{figure*}[t]
\begin{AIbox}{An example prompt for extracting objects from audio}
\footnotesize
I will give you a sentence. Please extract some information I need in a JSON format. Sentence: '{caption}'

My requirement: \newline
1. Extract instruments and return as a list \newline
2. Extract emotions and return as a list \newline
3. Extract music genres and return as a list \newline
4. Extract scenes and return as a list \newline
5. All words must be found in the sentence. \newline
6. Return a JSON format without any other words. \newline
7. Words must be extracted from the corresponding caption. \newline

The return format should only be like this:
\begin{verbatim}
    {
        "instrument": [], 
        "emotion": [],
        "music genre": [],
        "scene": []
    }
\end{verbatim}
\end{AIbox}
\caption{An example prompt for extracting objects from audio.}
\label{fig:Object_prompt}
\end{figure*}
The prompt for asking GPT-4o mini to obtain instruments, emotions, and music styles from audio is shown in Figure~\ref{fig:Object_prompt}.

\subsection{
Modal Information Check Prompt}
\label{app:modal_information_check_prompt}

\begin{figure*}[t]
    \begin{AIbox}{Shortened Prompt for Modal Integration Check}
        \scriptsize 
        \begin{verbatim}
"You analyze descriptions from audio. 
'final_cap' is a comprehensive summary.  
Identify source captions ('audio_caption', 'speech_caption', 
'music_caption', 'video_caption') essential for 'final_cap' 
using provided JSON data: {cap_str}.  

Requirements:  
1. List contributing caption types.  
2. Return as string keys list.  
3. Format: ['type1', 'type2']"
        \end{verbatim}
    \end{AIbox}
    \caption{Concise prompt for modal info check}
    \label{fig:modal_information_check_prompt}
\end{figure*}

The prompt we used to check if the modal information is used during the fusion is shown in Figure~\ref{fig:modal_information_check_prompt}.

\begin{figure*}[t]
\label{fig:clap case}
\begin{AIbox}{Examples for audios with different clap scores.}
\footnotesize

Here we show the severity of hallucinations in audio captions under different clap similarity intervals. The red - marked parts are the hallucinatory parts of the audio captions.

\end{AIbox}
\caption{Examples for audios with different clap scores.}
\end{figure*}

\begin{figure*}[t] 
\label{fig:mc_prompt}
\begin{AIbox}{An example prompt for multi-choice questions}
{\bf Prompt:} \\
Please provide a comprehensive video description focusing exclusively on observable visual elements, including timestamps:

**1. Key Entities \& Actions with Timestamps:** \\
- List main objects/subjects and their visible actions with approximate timestamps (MM:SS format)\\
- Describe: \\
  * Object/subject movements and interactions\\
  * Material properties (metal, wood, liquid)\\
  * Timing of significant visual events\\

**2. Scene Description with Timeline:**\\
- Overall scene dynamics and visual interactions\\
- Notable visual events with timestamps:\\
  * Object collisions or impacts\\
  * Movement patterns\\
  * Material changes\\
  * Human/animal visible actions\\
- Environmental context (indoor/outdoor, spatial relationships)\\

**3. Overall Description with Chronological Flow:**\\
- Provide a comprehensive visual narrative of the video\\
- Include timestamps for key moments and transitions (MM:SS format)\\
- Focus on observable actions, and movements\\
- Use specific, action-oriented language\\
- Present events in chronological order with clear time markers\\

Guidelines:\\
- Describe only directly visible elements\\
- Focus on observable actions and movements\\
- Note material properties and physical interactions\\
- Include **timestamps** for all significant events\\
- Timestamp **should not** exceed the duration of the video\\
- Use precise descriptive language for visual elements\\
- Avoid assumptions about non-visible elements\\
- Maintain strict focus on visual information\\

Example:\\
Instead of "A car's engine roars as it accelerates"\\
Write "00:01 - A red sports car with chrome detailing accelerates down a paved road, tires creating visible spray on wet asphalt"\\
"00:02 - The car's rear suspension compresses during acceleration, exhaust emitting visible vapor"\\
"00:03 - The car's engine roars as it accelerates"\\
\end{AIbox} 
\caption{An example prompt for multi-choice questions.}
\end{figure*}

\begin{figure*}[t]
\label{fig:integration_prompt_1}
\begin{AIbox}{Prompt for integration}
\footnotesize
{\bf Prompt:} \\
Rigorous Multimodal Information Integration and Purely Audio Description Expert\\

\textbf{Core Task}\\
You are an expert specializing in audio information processing. Your goal is to: integrate and analyze textual descriptions from multiple modalities as input, perform cross-referencing and correction while strictly controlling cross-modal information interference, and ultimately generate a description that is \textbf{purely about the audio content}, accurate, detailed, and fluently written in English, annotating potential ambiguities \textbf{based solely on auditory perception}. \textbf{It is strictly prohibited to include any visual information, specific speech dialogue content, or ambiguity annotations based on audio-visual inconsistencies in the final output.}\\

\textbf{Input Information Sources (May contain errors, hallucinations, or be incomplete)}
\begin{itemize}
    \item Audio Tags: A set of sound category tags annotated by humans, along with their corresponding quality estimations (confidence scores). Represents the most prominent human-perceived acoustic features in the audio. \textbf{These tags are highly reliable, especially those with high percentages}, but may not comprehensively cover all information in the audio. \textbf{The format is \texttt{TagName(Percentage\%)}. e.g., \texttt{Speech(100\%)}.} If empty, it indicates no human-annotated tag information is available.
    \item Audio Description: A textual description of the audio content (may include sound events, ambient sounds, music, vocal characteristics, etc.). This is an \textbf{important basis} for describing audio facts and needs to be cross-validated with tags and music descriptions.
    \item Speech Content: The textual result from Automatic Speech Recognition (ASR). \textbf{This information is used only to confirm the presence of human voice, determine general vocal characteristics (e.g., speech vs. non-linguistic sounds, presence of distinct emotions [non-content related]), and assist in inferring possible scenarios or event backgrounds. Its specific textual content (including paraphrasing or summarization) must never appear in the final output.} If empty, it indicates no distinct human voice, or other non-linguistic vocalizations (e.g., gasping, crying, background babble).
    \item Music Description: A description of musical elements (features, instruments, rhythm, etc.) and other sound scenes. \textbf{Music-related features herein are highly reliable. If empty, it indicates no distinct music.} Other non-music descriptions (e.g., environment, human voice) have lower priority and primarily depend on \texttt{"Audio Tags"}, \texttt{"Audio Description"}, and \texttt{"Speech Content"} for judgment.
    \item Video Description: A textual description of the video frames. \textbf{Used only under specific conditions} (see \texttt{"Active Correction"} in Processing Steps, step 2) \textbf{to actively assist in identifying auditorily ambiguous sound sources}, and \textbf{to identify inconsistencies with auditory information (this inconsistency is only an internal decision-making flag for the model, not used to generate the output ambiguity list)}. \textbf{Never} speculate or describe the source, location, or on-screen actions of sounds based on video information itself. If empty, it indicates a lack of visual auxiliary information.
\end{itemize}

\textbf{Processing Steps}\\
Please strictly follow the steps below:
\begin{enumerate}
    \item \textbf{Multimodal Information Parsing}: 
    \begin{itemize}
        \item Separately interpret each input description to extract core sound events, sound source characteristics, environmental ambiance, and musical elements.
        \item Specifically parse \texttt{"Audio Tags"} to extract tag names and their confidence scores.
        \item \textbf{Special Note}: From \texttt{"Speech Content"} (ASR results), primarily determine \textbf{if human voice is present} and its \textbf{non-content features}. In conjunction with its textual content (\textbf{used only for auxiliary understanding}), \textbf{assist in inferring} possible environments, emotional tones, or types of acoustic events, but \textbf{never judge} speaker gender, age, or other personal characteristics based on ASR content, and \textbf{never quote, paraphrase, or summarize} the specific textual content.
    \end{itemize}
\end{enumerate}
\end{AIbox}
\caption{Prompt for integration.}
\end{figure*}

\begin{figure*}[t]
\label{fig:integration_prompt_2}
\begin{AIbox}{Prompt for integration Cont.}
\footnotesize
\begin{enumerate}[start=2]
    \item \textbf{Auditory Fact Determination and Cross-modal Correction}: 
    \begin{itemize}
        \item \textbf{Initial Determination of Auditory Facts}: First, based on \texttt{"Audio Tags"} (\textbf{especially high-confidence tags, which have the highest priority for determining the types of sounds included in the tags}), \texttt{"Audio Description"}, \texttt{"Music Description"} (especially the music part), and \texttt{"Speech Content"} (presence of human voice and inferred characteristics), preliminarily determine auditorily perceived sound events, sound sources, ambient sounds, and music features. Identify and attempt to correct contradictions within these audio information sources (tags, audio description, music description, ASR inference), with the priority rule: \textbf{High-confidence \texttt{"Audio Tags"} > Music part of \texttt{"Music Description"} $\approx$ \texttt{"Audio Description"} > \texttt{"Speech Content"} (presence of human voice) > Low-confidence \texttt{"Audio Tags"} > Non-music part of \texttt{"Music Description"}}.
        \item \textbf{Cross-modal Validation and (Conditional) Active Correction (for video information):} After the initial determination of auditory facts, introduce \texttt{"Video Description"} for cross-validation. Its role is:
        \begin{itemize}
            \item \textbf{Active Correction (when audio information is ambiguous and video provides clear evidence):} If the initially determined auditory fact (based on audio information sources) describes a \textbf{general sound type that could have multiple auditory interpretations} (e.g., a rumbling sound, a clicking sound, a rustling sound), \textbf{and} the \texttt{"Video Description"} clearly shows an object or event that is \textbf{highly relevant to this general sound type and is a plausible sound source} (e.g., the video clearly shows an airplane making a rumbling sound, or a person clicking a mouse making a clicking sound, or clothes/fabric in motion making a rustling sound), then \textbf{the information provided by the video should be adopted to more precisely identify the general sound as a specific source or type} (correcting rumbling to airplane sound, clicking to mouse click, rustling to fabric rustle). \textbf{Note: If a high-confidence tag in \texttt{"Audio Tags"} already clearly indicates the specific sound type, then this sound is no longer considered a 'general sound type with multiple auditory interpretations,' and this active correction step no longer applies to this sound.} Under these limited and clear conditions, video information is used to \textbf{enhance} the understanding of audio facts, making the description more precise.
            \item \textbf{Identifying Inconsistencies or Lack of Corroboration (when video cannot clearly corroborate or conflicts):}
            \begin{itemize}
                \item If the sound event described by the initially determined auditory facts \textbf{does not have a clearly corresponding visual sound source} in the \texttt{"Video Description"}, or if the visual information \textbf{is inconsistent with or contradicts the perceived location or state of the sound source}, then \textbf{video information must never be used to negate or modify known auditory facts}. In such cases, the model should \textbf{internally flag} the presence of an audio-visual inconsistency or lack of visual corroboration. \textbf{This flag is only used in subsequent steps to adopt conservative wording when generating the final audio description and must never directly generate an ambiguity entry for output.}
                \item \textbf{It is strictly prohibited to speculate, describe, or alter judgments about the sound event itself based on video information that cannot corroborate the audio (e.g., hearing a rumbling sound, the video shows the sky, but one cannot speculate it's an airplane sound unless the video explicitly shows an airplane).}
            \end{itemize}
        \end{itemize}
        \item \textbf{Determine Corrected Auditory Facts}: Based on the results of the above multimodal cross-validation, determine the final auditory facts. The priority rule is listed above. \textbf{When cross-modal information conflicts, audio information sources conflict internally, or there is high uncertainty (especially a lack of high-confidence tags or clear video corroboration for audio) making it difficult to determine auditory facts, the determined facts should reflect \textbf{extreme} conservatism, \textbf{preferring to omit uncertain information rather than speculating based on non-auditory information.}} The model should internally retain a flag for the uncertain origin of audio information (e.g., whether it's due to a lack of high-confidence tags, lack of support from audio description, or lack of video corroboration), to generate appropriately conservative descriptions in step 5.
        \item \textbf{Emotion Inference and Correction}: If the emotion of a sound event (e.g., human voice, whose emotion can be inferred with ASR content assistance) conflicts with the emotion of background music, a comprehensive judgment must be made to provide the most likely primary emotional tone, but this is still based on auditory and ASR-assisted inference, without introducing visual information.
    \end{itemize}

    \item \textbf{Purely Auditory Ambiguity Reasoning and Annotation}: 
    \begin{itemize}
        \item \textbf{Focus Solely on Pure Audition}: Based on the \textbf{determined auditory facts} (which have considered tags and correction results), sound characteristics, common possibilities of auditory confusion, and potential auditory understanding biases in the original audio description, infer potential auditory understanding ambiguities that can be \textbf{perceived or reasonably inferred solely through hearing}.
    \end{itemize}
\end{enumerate}
\end{AIbox}
\caption{Prompt for integration Cont.}
\end{figure*}

\begin{figure*}[t]
\label{fig:integration_prompt_3}
\begin{AIbox}{Prompt for integration Cont.}
\footnotesize
\begin{enumerate}[start=4]
    \item 
    \begin{itemize} 
        \item \textbf{Sources of Ambiguity}:
        \begin{itemize}
            \item \textbf{Auditory Similarity or Vagueness of the Sound Itself}: Some sounds may be auditorily similar to others and easily confused (e.g., vehicle sound vs. airplane sound, typing sound vs. light tapping sound). The sound's own quality, distance, or reverberation can also lead to vagueness or difficulty in determining the source.
            \item \textbf{Polysemy of Auditory Association}: A sound event may reasonably correspond auditorily to multiple different sound sources or situations (e.g., a ``bang'' can have multiple causes, footsteps might come from multiple people).
            \item \textbf{Potential Purely Auditory Biases in the Original Audio Description}: If, after multimodal correction, the original \texttt{"Audio Description"} is found to have incorrect or imprecise judgments about sound events or sources (and this error/imprecision is not caused by audio-visual inconsistency but by potential misinterpretations of audition itself), one should infer what common purely auditory misinterpretations the original description might have been based on.
        \end{itemize}
        \item \textbf{Strictly Exclude Non-Auditory Information as a Source of Ambiguity}: Ambiguity annotation must \textbf{only} revolve around pure auditory perception and the associations arising therefrom. \textbf{It is absolutely not allowed} to use audio-visual synchronization, the way sound sources are presented on screen, or any visual content as the source or descriptive content of an ambiguity.
    \end{itemize}

    \item \textbf{Information Reliability Assessment and Final Output Decision}: 
    \begin{itemize}
        \item \textbf{In this step, based on the analysis and correction results from steps 1-3, comprehensively assess the reliability and completeness of the determined auditory facts. In particular, consider whether high-confidence audio tags support key sound events.}
        \item \textbf{If it is judged that the determined auditory facts are extremely scarce, various audio information sources (tags, audio description, music description, ASR inference) severely conflict and auditory facts cannot be reliably reconstructed, or even if tags exist but their confidence is generally very low and contradicts other information, the model directly outputs the unique specific string \texttt{UNCERTAIN\_AUDIO\_INFORMATION\_DETECTED}.}
        \item \textbf{Otherwise (if the determined auditory facts are sufficiently reliable and complete), proceed to the next step (generating JSON).}
    \end{itemize}

    \item \textbf{Generate Final Pure Audio Description (Audio Caption)}: 
    \begin{itemize}
        \item \textbf{Execute this step only after passing the reliability assessment in step 4.}
        \item \textbf{Pure Audio Focus}: Generate a fluent, accurate, detailed, and concise English audio description. \textbf{Describe only what can be heard and its purely auditory characteristics} (e.g., sound source type [prioritizing those confirmed by high-confidence tags or clearly identified through active video correction], nature of sound events, type of ambient sound, music features, non-content features of human voice, spatial sense, loudness, timbre, duration, rhythm, etc.).
        \item \textbf{Integration and Augmentation}: Integrate all valid auditory facts determined after multimodal correction (including those from audio tags, audio description, music description, ASR inference, and sound source types actively corrected via video). Supplement necessary auditory details of the scene (e.g., indoor/outdoor inferred from ambient sounds). \textbf{If the model has internally flagged uncertainty in the audio information} (e.g., lack of high-confidence audio tags supporting key sound events, original audio description being auditorily vague and lacking clear video corroboration, or internal conflicts within audio information sources), \textbf{the final description must reflect this uncertainty, but through \textbf{cautious wording} to describe \textbf{the perceived sound itself}, rather than directly stating the uncertainty or vagueness.} Use phrases like ``sounds like,'' ``appears to be,'' ``potentially,'' ``suggests,'' ``a sound resembling X is heard'' to express identification of less certain sound sources or events. \textbf{Crucially, avoid sentences that explicitly state an inability to determine something or that something is ambiguous (e.g., do not say ``the source cannot be determined'' or ``it is ambiguous whether X is present''). Instead, directly omit highly uncertain details or use cautious wording for what \textit{might} be perceived.}
        \item \textbf{Objective and Accurate}: Base inferences on determined auditory facts, avoiding subjective speculation and over-extension. The description content must be supported by input text. Prohibit the introduction of irrelevant ``new information,'' unless it is reliable auditory inference based on multiple audio information sources (e.g., inferring the scene from ambient sounds). Ensure the description integrates facts confirmed by high-confidence tags, but \textbf{never mention the confidence percentages themselves.}
    \end{itemize}
\end{enumerate}
\end{AIbox}
\caption{Prompt for integration Cont.}
\end{figure*}

\begin{figure*}[t]
\label{fig:integration_prompt_4}
\begin{AIbox}{Prompt for integration Cont.}
\setlist{itemsep=0pt, parsep=0pt, topsep=2pt, partopsep=0pt}
\begin{itemize}
        \item \textbf{Cultural/Emotional Cues}: If the sound contains clear cultural symbols or strong emotions, these can be briefly cued, but must be based on input audio evidence (e.g., emotion in human voice inferred from ASR, or emotion reflected by music features).
        \item \textbf{Final Check}: Ensure this description \textbf{absolutely contains no visual elements} (objects, colors, actions, visual scenes, etc.). Even if the sound source type has been determined through high-confidence tags or active video correction, \textbf{never describe the visual location, visual form, or specific on-screen behavior of that sound source}. \textbf{Absolutely prohibit} the output of any specific speech text content (quotation, paraphrase, summary).
\end{itemize}
\textbf{Output Format Requirements}\\
\textbf{For most cases (i.e., when passing the reliability assessment in step 4), please strictly generate structured English output in the following JSON format (without any other explanations). However, in the special case of ``scarce/unverifiable information'' defined in step 4 of the processing flow, the model should directly output the predefined string \texttt{UNCERTAIN\_AUDIO\_INFORMATION\_DETECTED} instead of JSON.}
\begin{verbatim}
    {
      "Potential ambiguities": [ // List potential ambiguities based purely on 
      auditory perception (English sentences). Does not include ambiguities 
      requiring visual information
      to understand, nor ambiguities based on audio-visual inconsistencies.
        "Ambiguity description 1 based solely on auditory perception.",
        "Ambiguity description 2 based solely on auditory perception.",
        ...
      ],
      "Audio caption": "Final audio description focusing solely on audible elements 
      and their auditory characteristics, detailed and fluent English. Use 
      conservative language when audio facts are uncertain based on internal 
      assessment." 
      // Final pure audio description (concise and clear English sentence)
    }
\end{verbatim}

\textbf{Key Considerations}:
\begin{itemize}
    \item Output Language: \textbf{English}.
    \item Ignore Empty Inputs: If a modal description is empty, ignore that information source.
    \item \textbf{Strictly Prohibited}: Including any visual information (objects, colors, actions, visual scenes, visual location/form/on-screen behavior of sound sources, audio-visual synchronization, etc.) in the final output (including \texttt{Audio caption} and \texttt{Potential ambiguities}). Even when dealing with audio-visual inconsistencies or unknown sound sources, never speculate, describe, or mention any visual content in the output.
    \item \textbf{Strictly Prohibited}: Including any specific speech text content (quotation, paraphrase, summary, etc.) in the final output. Speech information is only used to infer the presence of human voice, non-content features, and to assist in understanding the scene ambiance.
    \item Maintain Objectivity: Base inferences on determined auditory facts, avoiding subjective speculation and over-extension. Information not supported by the input or derived through reliable auditory inference must not appear in the output.
    \item When the model internally flags audio information as uncertain (even if \texttt{UNCERTAIN\_AUDIO\_INFORMATION\_DETECTED} is not triggered), the final \texttt{Audio caption} must strictly use cautious wording to describe the sound itself, focusing on auditory perception. \textbf{It is strictly prohibited to directly state uncertainty or ambiguity; only provide confirmed acoustic facts and do not mention uncertain content in the output.}
    \item \textbf{High-confidence audio tags are the highest priority source for determining sound type facts, but specific confidence values are not allowed in the output.}
\end{itemize}

\end{AIbox}
\caption{Prompt for integration Cont.}
\end{figure*}

\section{Dataset Samples}
\label{app:samples}
The code and dataset of this paper can be found in \url{https://github.com/satsuki2486441738/FusionAudio}.
\subsection{Case study}
Table~\ref{tab:crossmodal_examples} presents representative FusionAudio captions, annotated with their information sources. These examples illustrate FusionAudio's ability to synthesize and reason across modalities, generating descriptions that go beyond mere aggregation to provide holistic, context-rich interpretations.

\begin{table}[htbp!]
\centering
\caption{Example of FusionAudio caption generation with annotated information sources.}
\begin{tabular}{p{3.5cm}|p{10cm}}
\toprule

\textbf{Audio Caption} & 
The audio is characterized by the sound of brass instruments, possibly trumpets, playing music.  \\
\midrule
\textbf{Speech Caption} & 
(empty) \\
\midrule
\textbf{Music Caption} & 
This is an instrumental marching band piece. The brass section plays the melody, while bagpipes provide a counter-melody. The mood is optimistic and cheerful. This piece may be used as background music for a national patriotic event.\\
\midrule
\textbf{Video Caption} & 
Chronological overall description: \newline
- \textbf{00:00}: The video opens with close-ups of musicians, highlighting their focused expressions and the shiny brass instruments. The musicians are seated in a row, each holding a saxophone or trombone, preparing to play. \newline
- \textbf{00:01}: As the conductor steps forward facing the musicians, a blue-and-white striped shirt appears in the frame. The conductor's gestures begin to guide the musicians as the performance starts. \newline
- \textbf{00:02 - 00:10}: The camera remains relatively still, capturing the musicians playing their instruments. The conductor's hand movements synchronize with the music, guiding the ensemble. The musicians' fingers move rhythmically on the keys and valves, producing harmonious sounds. The conductor continues to lead, ensuring cohesion. \newline
- \textbf{00:07 - 00:10}: The camera pans slightly right, revealing more trombone players at the far right. The additional musicians add depth to the ensemble, enhancing the overall sound and visual impact. The video ends with the musicians continuing to play under the conductor's direction. \\
\midrule
\textbf{Reasoning Content} & 
\textbf{The audio caption mentions "trumpets," but the music and video captions indicate a brass ensemble with saxophones, trombones, and bagpipes. The music caption is more reliable for instrument identification, so the final description should prioritize saxophones, trombones, and bagpipes.} \newline

\textbf{There is no speech content, which matches the empty speech caption.} \newline

\textbf{The mood is cheerful and ceremonial, suitable for a patriotic event, as inferred from the music caption.} \newline

\textbf{Potential ambiguities include confusion about the specific types of brass instruments and the presence of bagpipes, as well as uncertainty about whether the performance is indoors or outdoors.} \newline

... (Further details omitted for brevity) \\
\midrule
\textbf{Final Caption} & 
A lively instrumental piece performed by a brass ensemble, with saxophones and trombones carrying the main melody and bagpipes providing a contrasting counter-melody. The music maintains the rhythmic precision characteristic of a formally conducted ensemble, featuring the harmonious and dynamic qualities typical of patriotic or ceremonial performances. No prominent vocal content. \\
\bottomrule
\end{tabular}
\label{tab:crossmodal_examples}
\end{table}

\subsection{Samples of different CLAP score}

\begin{table}[htbp]
\centering
\caption{The demonstration of the hallucination which is marked in red of audio captions within different clap similarity intervals}
\label{tab:clap_example}
\begin{tabularx}{\textwidth}{
  >{\centering\arraybackslash}m{\dimexpr0.2\textwidth-2\tabcolsep} 
  >{\centering\arraybackslash}m{2cm}                             
  >{\raggedright\arraybackslash}m{\dimexpr0.8\textwidth-2\tabcolsep-2cm} 
}
\toprule[1pt] 
\textbf{Clap Similarity Intervals} & \textbf{Audio ID} & \textbf{Caption} \\
\midrule[0.6pt] 

0.0-0.1 & -wyJ2cab4ic & A speech with strong tonal urgency is delivered, accompanied by \textcolor{red}{persistent breathing sounds} and \textcolor{red}{faint intermittent background} activity suggesting an indoor environment. The speaker's vocal cadence appears strained, potentially reflecting either \textcolor{red}{passionate delivery} or underlying emotional tension. \\
\midrule[0.6pt] 

0.1-0.2 & -4t1LMiiHp4 & A clear male speech is delivered with a strong vocal presence, accompanied by \textcolor{red}{dynamic acoustic drums}, a groovy bassline, and intermittent tambourine shakes in the background. Sporadic applause and crowd cheering weave \textcolor{red}{throughout the speech}, creating an energetic and engaged atmosphere. The musical elements maintain a steady rhythmic foundation while the vocal delivery appears \textcolor{red}{deliberate} and focused. \\
\midrule[0.6pt] 

0.2-0.3 & 04Q\_WeM7VIU & Continuous music with a groovy bass line, percussive drum patterns, keyboard harmonies, and synth brass melodies is heard in a lively setting. Intermittent male speech occurs in an upbeat tone, overlapping with the music's rhythmic elements. The recording exhibits \textcolor{red}{mono audio} and \textcolor{red}{background noise}, suggesting a live performance environment with frequent equipment adjustments and energetic vocal exchanges. \\
\midrule[0.6pt] 

0.3-0.4 & -CCsZneHL6s & A solo violin performs a slow, emotive melody with a smooth bowing technique, accompanied by steady rhythmic percussive sounds suggesting a handpan or similar instrument. The performance takes place in an indoor environment with \textcolor{red}{subtle background reverberation}, indicative of a studio or concert space. The audio quality is slightly degraded, but the interplay between the sustained violin tones and precise percussive elements creates a harmonious, intimate atmosphere. \\
\midrule[0.6pt] 

0.4-0.5 & -EKjvd8q\_A0 & The audio features a lively and energetic performance with rhythmic maracas, congas, and an accordion, accompanied by a saxophone adding depth. The upbeat tempo and festive soundscapes suggest a cultural celebration or live musical event. \\
\midrule[0.6pt] 

0.5-0.6 & 00Twebqicmo & The audio is dominated by powerful car engine revving and acceleration sounds, accompanied by continuous background music. The combination of loud mechanical noises and energetic musical accompaniment creates a high-intensity atmosphere characteristic of an automotive event. Intermittent engine echoes suggest open-air acoustics typical of a racetrack or exhibition setting. \\
\bottomrule[1pt] 
\end{tabularx}
\end{table}
\label{app:clap_score_samples}
Table~\ref{tab:clap_example} presents the hallucination situations of FusionAudio captions within different CLAP similarity intervals.

\subsection{Situations where multimodal contextual cues work}

Our multimodal approach is designed to excel in challenging audio understanding scenarios (Table~\ref{tab:captioning_scenarios}), such as interpreting audio in adverse conditions, achieving high-level semantic understanding (e.g., nuanced music interpretation), and enabling fine-grained acoustic entity recognition. Addressing these scenarios highlights the benefits of comprehensive multimodal integration.

\begin{table*}[t] 
\centering
\small
\renewcommand{\arraystretch}{1.3} 
\caption{Key use-case scenarios where integrating multimodal contextual cues can significantly improve audio captioning. Challenges are listed per sub-scenario. Representative datasets and samples are detailed in Appendix~\ref{app:subcategory_samples}.}
\begin{tabularx}{\textwidth}{
>{\centering\arraybackslash}p{3.2cm} 
>{\centering\arraybackslash}p{3.1cm} 
>{\raggedright\arraybackslash}X      
}
\toprule
\textbf{Scenario} & \textbf{Sub-Scenario} & \textbf{Key Challenges}   \\
\midrule
\multirow{2}{=}{\makecell[cc]{\\Adverse Acoustic\\Conditions}} 
    & \makecell[cc]{Scene Recognition in\\Complex Soundscapes}
    & \makecell[cl]{High inherent acoustic complexity; Interwoven\\multi-source information; Background noise}
\\
\cmidrule(lr){2-3}\addlinespace[1pt]
    & \makecell[cc]{Acoustically\\ Degraded Conditions}
    & \makecell[cl]{Recording device limitations; Synthetic Artificial \\noise interference}
\\
\midrule
\multirow{2}{=}{\makecell[cc]{High-Level\\Semantic\\Understanding}}
    & \makecell[cc]{Music Understanding}
    & \makecell[cl]{Musical Genre Analysis; Emotional Expression;\\ Artistic Intent; Aural Narratives}
\\
\cmidrule(lr){2-3}\addlinespace[1pt]
    & \makecell[cc]{Sound Understanding}
    & \makecell[cl]{Sound Implied Information; Attributes Inference}
\\
\midrule
\makecell[cc]{Fine-grained\\Information Recognition}
    & \makecell[cc]{Acoustic Entity\\Recognition}
    & \makecell[cl]{Subtle acoustic cue discernment}
\\
\bottomrule
\end{tabularx}
\label{tab:captioning_scenarios}
\end{table*}

\subsection{Samples of different sub-scenario}
\label{app:subcategory_samples}

\begin{table}[htbp]
\centering
\caption{Dataset and examples corresponding to each sub-scenario, where cls is the classification task}
\renewcommand{\arraystretch}{1.3} 
\begin{tabularx}{\textwidth}{
>{\centering\arraybackslash\hsize=0.8\hsize}X
>{\raggedright\arraybackslash\hsize=1.1\hsize}X
>{\raggedright\arraybackslash\hsize=1.1\hsize}X 
}
\toprule
\textbf{Sub-Scenario} & \textbf{Datasets(quantity)} & \textbf{Examples}   \\
\midrule
     \makecell[cc]{ Scene Recognition in\\Complex Soundscapes}
    & \makecell[cl]{AIR-Bench:\\~~~~Acoustic scene cls(2,000) \\ UrbanSound8K(8,732)}
    & \makecell[cl]{\textit{Identifying child playing scene} \\ \textit{Identifying kitchen scene}}
\\
\midrule 
     \makecell[cc]{Acoustically Degraded\\Conditions}
    & \makecell[cl]{TAU Urban Sound-Mobile(5,265)\\FSDnoisy18K(947)}
    & \makecell[cl]{\textit{Identifying street pedestrian sound} \\ \textit{Identifying metro station scene}}
\\
\midrule
\midrule 
     \makecell[cc]{Music Understanding}
    & \makecell[cl]{AIR-Bench:\\~~~~Genre cls(2,000)\\~~~~MusicAQA(814)\\~~~~Mood detection(2,000)\\~~~~Chat-Music(500)}
    & \makecell[cl]{\textit{Identifying music genre} \\ \textit{Character portrayed by the tune}\\ \textit{Trumpet\&accordion's role in texture}}
\\
\midrule 
     \makecell[cc]{Sound Understanding}
    & \makecell[cl]{AIR-Bench:\\~~~~SoundAQA(2,000)\\~~~~Chat-Sound(500)\\AudioBench:\\~~~~Audio-Scene QA(9,131)}
    & \makecell[cl]{\textit{Location of dripping water} \\ \textit{Possible actions with the liquid}\\ \textit{Indications of a busy road}}
\\
\midrule 
\midrule 
     \makecell[cc]{Acoustic Entity\\Recognition}
    & \makecell[cl]{AIR-bench:\\~~~~Vocal sound cls(1,000)\\~~~~Music instruments cls(2,000)\\ESC-50(2,000)\\FSD50K(10,231)}
    & \makecell[cl]{\textit{Instrument recognition} \\ \textit{Acoustic event/ontology recognition}\\ \textit{Acoustic scene type recognition}}
\\
\bottomrule 
\end{tabularx}
\label{tab:captioning_scenarios_tasks}
\end{table}

Table~\ref{tab:captioning_scenarios_tasks} shows the example dataset for each sub-scenario and corresponding example samples.

\section{More on Dataset Statistics}

\subsection{Embedding Space Quantitave Analysis}
\label{app:qutatitave_embedding}
Table~\ref{tab:distances_combined_cols} presents a comprehensive comparison of inter- and intra-category embedding distances across different datasets. The analysis focuses on three key audio categories: Music (M), Vehicle (V), and Speech (S). Our proposed FusionAudio dataset demonstrates superior performance across all metrics. For inter-category distances, where higher values indicate better category separation, FusionAudio achieves significantly larger distances between different audio types (M-V: 0.7230, M-S: 0.5369, V-S: 0.5943) compared to competing datasets. This indicates that our dataset enables models to learn more discriminative representations that effectively distinguish between different audio categories. Simultaneously, FusionAudio exhibits smaller intra-category distances (Music: 0.8084, Vehicle: 0.7406, Speech: 0.8204), reflecting greater consistency within each category. The substantial improvement in both metrics—maximizing inter-category separation while minimizing intra-category variation—confirms that FusionAudio produces more cohesive and well-structured embedding spaces. This balance is crucial for downstream tasks such as audio classification, retrieval, and generation, as it facilitates more accurate identification and characterization of audio content while maintaining the nuanced variations within categories.

\begin{table}[htbp!]
  \centering
  \caption{Inter- (M-V, M-S, V-S) and Intra- (M, V, S) category embedding distances. Best inter-distances (higher) and intra-distances (lower) are bolded.}
  \label{tab:distances_combined_cols}
  \sisetup{
    round-mode=places,
    round-precision=4,
    table-format=1.4,
    detect-weight,
    mode=text
  }
  \begin{tabular}{l S S S S S S}
    \toprule
    \multirow{2}{*}{\small \textbf{Dataset/Method}} & \multicolumn{3}{c}{Inter-category distance ↑} & \multicolumn{3}{c}{Intra-category distance ↓} \\
    \cmidrule(lr){2-4} \cmidrule(lr){5-7} 
                   & {M -- V} & {M -- S} & {V -- S} & {Music} & {Vehicle} & {Speech} \\
    \midrule
    FusionAudio           & \textbf{0.7230} & \textbf{0.5369} & \textbf{0.5943} & \textbf{0.8084} & \textbf{0.7406}          & \textbf{0.8204} \\
    ASC            & 0.5685 & 0.4137 & 0.4523 & 0.8638 & 0.8216          & 0.8724 \\
    Auto-ACD       & 0.5685 & 0.4137 & 0.4523 & 0.8645 & 0.8402          & 0.8915 \\
    Sound-VECaps   & 0.5232 & 0.3770 & 0.4664 & 0.8578 & 0.7798 & 0.8920 \\
    \bottomrule
    \label{tab:distances_comparison_enhanced} 
  \end{tabular}
\end{table}

\end{document}